\documentclass[12pt,letterpaper,usenames,dvipsnames]{article}
\usepackage{jheppub}

\allowdisplaybreaks[2]  
\usepackage{bm} 
\usepackage{mathrsfs}   
\usepackage{dsfont}  
\usepackage{ytableau}   
       
\usepackage{float}
\usepackage{hhline}
\usepackage{mathtools}
\usepackage{hyperref}

\graphicspath{{figures/}}	 
\usepackage{tikz}
\usetikzlibrary{arrows,snakes,shapes.arrows,decorations.markings}
     \tikzset{>=triangle 45}
     \tikzstyle{gr}=[draw,circle,green!50!black,fill=green!50!black,scale=.6]
     \tikzstyle{Bl}=[draw,circle,blue,scale=.7]
     \tikzstyle{R}=[draw,circle,fill=red,scale=.7]
     \tikzstyle{bl}=[draw,circle,fill=black,scale=.2]
     \tikzstyle{bbc}=[draw,circle,fill=black,scale=.75]
 
\usepackage{array} 
\usepackage{multirow} 
\usepackage{colortbl} 
\usepackage{arydshln} 
\usepackage{stfloats}  

\usepackage{xcolor}   

\usepackage{url} 
\usepackage{verbatim} 

\def\bar{\overline}
\def\til{\widetilde}
\def\hat{\widehat}
\def\del{{\partial}}
\def\delb{{\bar\del}}
 
\def\vev#1{{\langle{#1}\rangle}} 
\newcommand{\beq}{\begin{equation}}
\newcommand{\eeq}{\end{equation}}
\newcommand{\bpmat}{\begin{pmatrix}}
\newcommand{\epmat}{\end{pmatrix}}
\newcommand{\bsmat}{\begin{smallmatrix}}
\newcommand{\esmat}{\end{smallmatrix}}
 
\def\v{\vee}
\def\^{\wedge}
\def\I{\mathds{1}}
\def\Tr{{\rm\, Tr}}
\def\Im{{\rm\, Im}}
\def\Re{{\rm\, Re}}
\def\U{{\rm\, U}}
\def\SU{{\rm\, SU}}

\def\SO{{\rm\, SO}}
\def\SL{{\rm\, SL}}
\def\GL{{\rm\, GL}} 
\def\Sp{{\rm\, Sp}}

\def\C{\mathbb{C}} 
\def\tCs{\til{\C^*}}

\def\I{\mathbb{I}}
\def\N{\mathbb{N}} 
\def\P{\mathbb{P}}
\def\Q{\mathbb{Q}} 
\def\R{\mathbb{R}} 
\def\Z{\mathbb{Z}} 

\def\gf{{\mathfrak g}}

\def\be{{\bf e}}
\def\bof{{\bf f}}

\def\bm{{\bf m}}

\def\bp{{\bf p}}

\def\bq{{\bf q}}

\def\tQ{{\til Q}}
\def\bR{{\bf R}}
\def\bs{{\bf s}}

\def\bu{{\bf u}}
\def\ub{{\bar u}}
\def\bv{{\bf v}}
\def\vb{{\bar v}}

\def\bw{{\bf w}}

\def\cC{{\mathcal C}}

\def\cE{{\mathcal E}}

\def\cM{{\mathcal M}}
\def\cN{{\mathcal N}}

\def\cP{{\mathcal P}}

\def\cS{{\mathcal S}}
\def\cT{{\mathcal T}}
\def\cV{{\mathcal V}}
\def\cW{{\mathcal W}}

\def\a{{\alpha}}

\def\bal{{\boldsymbol\a}}

\def\b{{\beta}}
\def\g{{\gamma}}
\def\G{{\Gamma}}
\def\d{{\delta}}
\def\D{{\Delta}}
\def\e{{\epsilon}}

\def\th{{\theta}}

\def\l{{\lambda}}
\def\L{{\Lambda}}
\def\m{{\mu}}
\def\bmu{{\boldsymbol\m}}
\def\n{{\nu}}

\def\r{{\rho}}

\def\s{{\sigma}}
\def\sb{{\bar\s}}
\def\sh{{\hat\s}}

\def\t{{\tau}}

\def\f{{\phi}}
\def\F{{\Phi}}
\def\vf{{\varphi}}

\def\w{{\omega}}
\def\mtM{{\mathscr M}}
\def\SpDtZ{{\Sp_\D(4,\Z)}}
\def\SpDrZ{{\Sp_\D(2r,\Z)}}

\title{The Singularity Structure of Scale-Invariant Rank-2 Coulomb Branches}
\author[1,2]{Philip C. Argyres,}
\author[3]{Cody Long,}
\author[4]{and Mario Martone}
\affiliation[1]{University of Cincinnati,
Physics Department, PO Box 210011, Cincinnati OH 45221}
\affiliation[2]{California Institute of Technology, Walter Burke Institute for Theoretical Physics, Pasadena CA 91125}
\affiliation[3]{Northeastern University, 
Department of Physics, Boston, MA 02115}
\affiliation[4]{University of Texas, Austin,
Physics Department, Austin TX 78712}
\emailAdd{philip.argyres@gmail.com}
\emailAdd{co.long@northeastern.edu}
\emailAdd{mariomartone@utexas.edu}

\abstract{
We compute the spectrum of scaling dimensions of Coulomb branch operators in 4d rank-2 $\cN{=}2$ superconformal field theories.  Only a finite rational set of scaling dimensions is allowed.  It is determined by using information about the global topology of the locus of metric singularities on the Coulomb branch, the special K\"ahler geometry near those singularities, and electric-magnetic duality monodromies along orbits of the $\U(1)_R$ symmetry.  
A set of novel topological and geometric results are developed which promise to be useful for the study and classification of Coulomb branch geometries at all ranks.
}
 
\begin{document}
\maketitle

\section{Introduction and summary of results}

A striking prediction from the study of the geometry of Coulomb branches (CBs) of 4d $\cN=2$ superconformal field theories (SCFTs) \cite{sw1, sw2, Argyres:1995xn, Minahan:1996fg, Minahan:1996cj} is that the spectrum of scaling dimensions of the CB operators of rank-1 theories can take only one of eight rational values.  This fact can be understood in terms of simple considerations involving the topology of the locus of metric singularities on the CB, positivity of the special K\"ahler metric on the CB, and the electric-magnetic (EM) duality monodromies around the singularities. In the rank-1 case, where the CB is 1 complex dimensional, the argument is particularly simple, because the metric singularity is a single point and all other points on the CB are related by the action of the spontaneously broken dilatation and $\U(1)_R$ symmetries.  The answer turns out to be closely related to Kodaira's classification of degenerations of elliptic fibers of elliptic surfaces \cite{KodairaI, KodairaII}.

In this paper we will generalize this argument to the rank-2 case, where the CB is 2 complex dimensional. The metric singularities now become a collection of complex curves, which are particular orbits of the combined holomorphic action of the dilatation and $\U(1)_R$ symmetries of the microscopic SCFT.  The EM duality monodromies around these singularities form a representation of the fundamental group of the non-singular part of the CB in $\SpDtZ$, which is the EM duality group.  The fundamental group of the CB turns out to be a knot group of torus links.  In addition, the special K\"ahler (SK) metric on the CB satisfies an integrability condition which was trivially satisfied in the rank-1 case, and so the topological, algebraic, and geometric ingredients in the rank-2 case are considerably more intricate than in the rank-1 case.  It may be worth noting that the analog for the rank-2 case of Kodaira's classification of singular elliptic fibers is the quite complicated classification \cite{Namikawa:1973} of singular genus-2 hyperelliptic fibers; however, this classification is only over a 1-dimensional base and does not incorporate any of the SK constraints, and is therefore insufficient for our purposes.

We will show that, at least to compute the spectrum of CB operator dimensions, one can bypass most of the intricacies of topology and details of $\SpDtZ$ conjugacy classes.  The key is to recognize that EM duality monodromies around cycles which are $\U(1)_R$ orbits have special properties.  In particular, the \emph{SK section}, i.e. the set of special coordinates and dual special coordinates, lies in an eigenspace of these monodromies, which includes a lagrangian subspace of the space of electric and magnetic charges, and the associated eigenvalues have unit norm.  This, together with a determination of the finite list of possible characteristic polynomials of the relevant EM duality monodromies, restricts the set of allowed CB dimensions to rational numbers satisfying some simple equations.   Furthermore, this set is finite if it is assumed that all CB dimensions are greater than or equal to 1.  This latter assumption follows from unitarity 
if the CB coordinates are vevs of CB chiral operators in the SCFT, a sufficient condition for which is that the CB chiral ring is freely generated \cite{Argyres:2017tmj}.

The resulting list of 24 allowed rank-2 CB scaling dimensions is given in Table \ref{ScaDimList}.  The dimensions greater than one range from $12/11$ to $12$, and, of course, the list includes the 8 rank-1 scaling dimensions. 

In addition to this concrete result on the spectrum of CB scaling dimensions, we develop a set of tools which will be useful for constructing all possible scale invariant rank-2 CB geometries.  Our key results are:  the algebraic description of the possible varieties, $\cV$, of CB singularities in \eqref{singularity}; the computation of the possible topologies of $\cV$ given in \eqref{L(1K1)}; the factorized description of the local EM duality monodromy linking components of $\cV$ in terms of $\Sp(2,\Z)\cong\SL(2,\Z)$ matrices given in \eqref{factorize}; the fact that the SK section is an eigenvector of $\U(1)_R$ monodromies with unit-norm eigenvalue \eqref{U1RMeigen}; the lagrangian eigenspace property \eqref{Meigenspace} and fact that all eigenvalues have unit norm \eqref{Meigenvalues} of the generic (knotted) $\U(1)_R$ monodromy; and the interrelations of the three topologically distinct $\U(1)_R$ monodromies recorded in Tables \ref{ScaDim}--\ref{ScaDim2}.

It may be helpful to put what we do here in the broader context of the program of systematically classifying CB geometries initiated in \cite{paper1, paper2, allm1602, am1604, Argyres:2016ccharges} for the rank-1 case.   At its core, this program relies on a two step process, each one in principle generalizable to rank-$r$ theories:
\begin{itemize}
\item[(i)] Classify the complex spaces that can be interpreted as CBs of SCFTs.   These are metrically singular spaces which are SK at their regular points ,and which have a well-defined action of the microscopic $\cN=2$ superconformal symmetry algebra.
\item[(ii)]  Further classify the possible mass or other relevant deformations of the set of geometries obtained in step (i).  These are complex deformations preserving an SK structure and satisfying various other physical consistency requirements, described in \cite{paper1}.
\end{itemize}
This paper presents first results in the rank-2 case towards realizing step (i).  We emphasize that finding the spectrum of rank-2 CB dimensions is not by itself a classification of scale-invariant rank-2 CB geometries.  For instance, despite the finiteness of the list of allowed scaling dimensions, it is not obvious that the set of distinct scale-invariant geometries is finite.  We do not attempt to address step (ii), the analysis of deformations, which seems considerably more challenging than step (i). 

Looking beyond rank-2, we note that it is possible to generalize many of the arguments in this paper to arbitrary rank $\cN=2$ SCFTs \cite{Argyres:2018urp}.  In particular these generalizations can be used to show that all the CB operators of $\cN=2$ SCFTs have rational scaling dimensions and, for a given rank, only a finite and computable set of possibilities is allowed.

The outline of the rest of the paper is as follows: Section \ref{topo} analyzes the topology of the set of singularities in the CB.  We denote the CB by $\cC$, and its subset of metrically singular points by $\cV$.  The set of metrically regular points, $\cM = \cC\setminus\cV$, is a 2-dimensional SK manifold.  After a brief review of the essential elements of SK geometry, we motivate some regularity assumptions which amount to assuming that $\cC\simeq\C^2$ as a complex space, and that $\cV$ does not have accumulation points in transverse directions.  We then introduce the holomorphic $\tCs$ action on $\cC$ induced by dilatations and $\U(1)_R$ transformations of the underlying SCFT.  We conclude Section \ref{topo} with the analysis of the topology of $\cV$, which can be the finite union of arbitrarily many $\tCs$ orbits, by computing the fundamental group of $\cM$.

Section \ref{phys} illustrates the arguments of Section \ref{topo} by analyzing examples of the simple case of rank-2 lagrangian SCFTs.  In particular, we show how to work out the topological structure of $\cV$ in these cases from familiar physical considerations.

Section \ref{secSKmetric} is concerned with the connection between the topology of the singularity locus $\cV$ and the EM duality monodromies around various cycles linking $\cV$.  This connection is forged by the SK geometry of $\cM$.  The central role is played by $\s$, the \emph{SK section}, which is the 4-component vector of special coordinates and dual special coordinates varying holomorphically on $\cM,$\footnote{Integer linear combinations of its components give the $\cN=2$ central charges in various low energy $\U(1)^2$ gauge charge sectors.}  and suffering EM duality monodromies around $\cV$. We start by showing that regularity of the SK metric on $\cM$ and the SK integrability condition imply that derivatives of $\s$ span a lagrangian subspace of the charge space.  We then argue that $\s$ has a well-defined finite limit almost everywhere on $\cV$, and that locally only two of its components can vanish identically along $\cV$.  This implies that the EM duality monodromy around a small circle linking a component of $\cV$ must have a simple factorized form in terms of $\Sp(2,\Z)$ monodromies, and allows us to complete an argument, started in Section \ref{topo}, showing that the scaling dimensions of the two CB coordinates are commensurate.  With commensurate coordinates, the orbits of the $\U(1)_R$ action on the CB are closed, and $\s$ is an eigenvector with a unit-norm eigenvalue of the EM duality monodromy around such orbits.  Furthermore, for a generic such orbit, the eigenspace in which $\s$ takes values is shown to contain a lagrangian subspace of the charge space.  These somewhat technical-sounding results provide a tight set of relations between the topology of $\cV$, its associated monodromies, and the scaling action on the CB.

Section \ref{monodromies} applies the results of Section \ref{secSKmetric} to derive the main result of the paper:  the full list of possible scaling dimensions of Coulomb branch operators of scale invariant rank-2 theories, collected in Table \ref{ScaDimList}, and a set of correlations among the conjugacy classes of the three different types of $\U(1)_R$ monodromies, recorded in Tables \ref{ScaDim}--\ref{ScaDim2}.  To derive the latter results some detailed information about the conjugacy classes of $\SpDtZ$ is used.  We conclude in Section \ref{conclu} with a summary of the likely next steps required in pursuit of constructing all scale-invariant rank-2 CB geometries.

The paper is completed by four appendices collecting both some known and some original technical results.  Appendix \ref{appRk1} reviews the construction of rank-1 scale invariant geometries, which we aim to generalize.  Though we do not strictly need it for any of the arguments of this paper, in appendix \ref{appanalytic} we derive the analytic form of the SK section in the vicinity of a point of $\cV\setminus\{0\}$ in terms of the Jordan block decomposition of the monodromy matrix around $\cV$.  Its explicitness may be helpful for making the reader's understanding more concrete.  Appendix \ref{appB0} collects some useful results about conjugacy classes of $\Sp(4,\R)$, reviewing generalized eigenspaces and some symplectic linear algebra along the way.  Finally, appendix \ref{appE0} describes the EM duality group, $\SpDtZ$, and derives the possible characteristic polynomials of their elements with only unit-norm eigenvalues.  Some elementary properties of cyclotomic polynomials are reviewed there as well.

\section{Topology of Coulomb branch singularities for rank-2 SCFTs}\label{topo}

In this section we will describe the topology of the set of metric singularities $\cV$ in a rank-2 CB $\cC$.  The metrically-regular points of the CB, $\cM := \cC \setminus \cV$, form a special K\"ahler (SK) manifold, which we assume to be 2-complex-dimensional.  In Section \ref{appSK} we review the essential elements of SK geometry.

In general how, or even whether, the complex structure of $\cM$ extends to $\cC$ is not clear from physical first principles.  In this paper we will therefore make the simplifying assumption that the complex (but not metric) structure of $\cM$ extends smoothly through $\cC$ (this assumption has physical implications, which are discussed below). Together with the assumption that the microscopic field theory is a SCFT, this will show that as a complex manifold, $\cC=\C^2$.
Also, as we explain in Section \ref{sec2.1}, we do not know how to rule out, from first principles, sets of metric singularities $\cV$ which are dense in $\cC$, and so we also assume that $\cV$ has no such accumulation points.

In Section \ref{cplxscaleact} we describe the holomorphic $\tCs$ action
of the combined (spontaneously broken) dilatation and $\U(1)_R$ symmetries on the CB of a SCFT.  We then classify the $\tCs$ orbits of points in $\cC$, which in our rank-2 case coincide with possible irreducible components of $\cV$.  In the case that a certain class of ``knotted" orbits occur as components of $\cV$, we deduce that the scaling dimensions of the CB operators must be commensurate.

In Section \ref{sec2.3} we describe the topology of $\cV$ in more detail.  Specifically, we compute $\pi_1(\cM)$ explicitly in terms of a simple set of generators and relations, using the results of a recent knot group computation \cite{Argyres:2019kpy}. To see the connection to knot groups (which are the fundamental groups of the complements of knots in $S^3$), note that by dilatation invariance it is enough to consider $X_\r:=\cV\cap S^3_\r$, where $S^3_\r$ is a three sphere of radius $\r$ centered at the origin of $\cC=\C^2$.  Then $X_\r$ is a deformation retract of $\cV$, which is a 1-real-dimensional manifold embedded in the 3-sphere --- i.e., a knot --- and $\pi_1(\cM)$ is the knot group of this knot.  We show that $X_\r$ is a \emph{torus link} --- a real curve which wraps a torus, $T^2$, $p$ times around one cycle and $q$ times around the other, with $\ell$ parallel and disconnected components.  Unknotted circles, wrapping $\ell_0$ times around the inside and $\ell_{\infty}$ times the outside of the torus, are allowed as well.  Examples of such $X_\r$'s are shown in Figures \ref{knot1}, \ref{knot2} and \ref{knot3}.

The importance of $\pi_1(\cM)$ is that the main arithmetic constraint on the SK geometry of $\cC$ arises from the fact that the EM duality monodromies of $\cC$ form a representation of $\pi_1(\cM)$.  The other main constraint is a geometric one, arising from the existence of an SK metric on $\cC$, and will be discussed in Section \ref{secSKmetric}.  These are the ingredients needed for constructing all rank-2 SCFT CB geometries via analytic continuation, generalizing the rank-1 classification.

\subsection{Basic ingredients of SK geometry}\label{appSK}

On the CB $\cC$ of vacua of a rank-$r$ 4d $\cN=2$ SUSY QFT, the manifold of generic points $\cM\subset\cC$ is described by a free $\cN=2$ $\U(1)^r$ gauge theory in the IR.  In particular, in this continuous set of vacua all fields charged with respect to the $r$ massless vector multiplets are massive.  Combinations of the vevs of the complex scalars of the $\U(1)$ vector multiplets are good complex coordinates on $\cM$, and the kinetic terms of the scalars define a K\"ahler metric on $\cM$.  Low energy $\cN=2$ supersymmetry implies the existence of an SK structure on $\cM$, which relates adjoint-valued (i.e., neutral) scalars to the $\U(1)$ vector fields.  The main ingredients are the charge lattice and its Dirac pairing, and the $\cN=2$ central charges, in terms of which the SK geometry of $\cM$ is completely determined.  There are various other formulations of SK geometry;  a paper that describes the main formulations, and is explicit about the equivalence of the various formulations, is \cite{Freed:1997dp}.

The charge lattice is a rank $2r$ lattice, $\Z^{2r}$, of the electric and magnetic $\U(1)^r$ charges of the states in the theory, along with the Dirac pairing $\vev{\bp,\bq} \in \Z$ for charge vectors $\bp, \bq \in \Z^{2r}$.  The Dirac pairing is non-degenerate, integral, and skew bilinear.  
The electric-magnetic (EM) duality group is the subgroup of the group of charge lattice basis changes, $\SpDrZ\subset \GL(2r,\Z)$ which preserves the Dirac pairing.\footnote{The reason for the $\SpDrZ$ notation is that we are allowing more general Dirac pairings than the canonical ``principally polarized" one.  This generality is important, for instance, if one wants to describe ``relative" field theories which appear naturally in first principles \cite{am1604,Regalado:2017} and class-$\cS$ \cite{Distler:2017} constructions of $\cN=2$ field theories.   $\SpDrZ$ is discussed in appendix \ref{appE0}, but since the facts that $\SpDrZ\subset\Sp(2r,\R)$ and that $\SpDrZ\subset\GL(2r,\Z)$ are the only facts we will use about $\SpDrZ$ in this paper, the distinction between $\SpDrZ$ and the more familiar $\Sp(2r,\Z)$ EM duality groups will not play any role.}  It is convenient to introduce a complex ``charge space" $V:=\C\otimes\Z^{2r} \simeq \C^{2r}$, and to extend $\vev{\cdot,\cdot}$ to $V$ by linearity.  

The central charge is encoded as a holomorphic section $\s$ of a rank $2r$ complex vector bundle $\pi: \cE\to\cM$ with fibers $V^*$ (the linear dual of the charge space) and structure group $\SpDrZ$.  We will call $\s$ the ``SK section"; its $2r$ complex components can be thought of as the $r$ special coordinates and $r$ dual special coordinates on $\cM$.  $V^*$ inherits a Dirac pairing and $\SpDrZ$ action from that on $V$.  The SK section is not unique: two sections $\s$ and $\s'$ related by $\s'=M^T\s$ for $M\in\SpDrZ$ define the same special K\"ahler geometry $\cM$.  

The SK section satisfies a further condition, which we will call the SK integrability condition:
\begin{align}\label{SKintegrability}
\vev{d\s\, \overset{\^}{,}\, d\s}=0\, ,
\end{align}
where $d$ is the exterior derivative on $\cM$.\footnote{In a basis of $V^*$ in which the Dirac pairing is given by the canonical symplectic form $J = (\bsmat 0&-1\\1&0\esmat)\otimes I_r$, then \eqref{SKintegrability} is equivalent to $\t^T=\t$ where $\t = BA^{-1}$ for $A$, $B$ the $r\times r$ matrices $A^i_j := \del \s^i/\del u^j$ and $B^i_j := \del \s^{r+i}/\del u^j$, where $u^j$ are complex coordinates on $\cM$.  $\t_{ij}$ is the complex $r\times r$ matrix of $\U(1)^r$ gauge couplings and theta angles.}  Some  consequences of this condition will be explored in Section \ref{secSKmetric} below.

The BPS mass of a dyon with charge vector $\bp$ is $|Z_\bp|$, where
\begin{align}\label{centralcharge}
Z_\bp:=\bp^T\s\, ,
\end{align}
is the central charge.  Here $\bp^T\s$ denotes the dual pairing $V\times V^* \to \C$.

The SK section also determines the K\"ahler geometry of $\cM$.  For instance, the K\"ahler potential on $\cM$ is given by
\begin{align}\label{Kpot}
K = i \vev{\bar\s, \s}\, ,
\end{align}
from which the metric can be readily obtained.  The consequences of demanding regularity of the K\"ahler metric on $\cM$ will be discussed in Section \ref{secSKmetric}.

Finally, the condition that $\s$ be a holomorphic section of $\cE$, and that $\cE$ has structure group $\SpDrZ$, simply means that $\s$ is a holomorphic vector field locally on $\cM$, and that the analytic continuation of $\s$ along any closed path $\g$ in $\cM$ will give a monodromy, $\s \xrightarrow[]{\,\, \g \,\,\,} M_\g^T \s$, with $M_\g \in \SpDrZ$.  By continuity, and since $\SpDrZ$ is discrete, if $\g$ is trivial in $\pi_1(\cM)$, then $M_\g = I$. Thus the monodromies $M_\g = M_{[\g]}$ only depend on the homotopy class $[\g]$ of $\g$, and $M_{[\g]}$ give a representation of $\pi_1(\cM)$ in $\SpDrZ$.

\subsection{Some regularity assumptions}
\label{sec2.1}

The CB $\cC$ is the metric completion of the SK manifold $\cM$ whose points correspond to vacua with $r$ massless vector multiplets and a mass gap for all other fields charged under the low energy $\U(1)^r$ gauge group. We will call the points of $\cC\setminus\cM$ --- which, by definition, are at a finite distance in the metric on $\cM$ --- the singularities of the CB, and denote the set of all singular points by $\cV \subset \cC$.  These are the vacua where some states charged under the $\U(1)^r$ gauge group become massless.  Note that $\cC$ need not be singular as a complex space at $\cV$; however, it will have metric singularities (non-smooth or divergent metric invariants) at all points of $\cV$, reflecting the breakdown of the  description of the low energy effective action in terms of free vector multiplets.

In fact, in general it is not obvious that $\cC$ need even inherit a complex structure at all. Even if $\cC$ is assumed to be a complex analytic space, such spaces can be quite complicated.  We propose to bypass possible ``strange'' behaviors by assuming:
\begin{align}\label{freeCBring}
\text{\emph{The complex structure of $\cM$ extends through $\cV$ to give a complex \emph{manifold} $\cC$.}}
\end{align}
This is certainly a stronger assumption than is needed to perform the following analysis; a discussion of weaker assumption will appear elsewhere \cite{Argyres:2018urp}.  In the case of a SCFT, this assumption has a clear physical interpretation: it implies that the (reduced) CB chiral ring of the SCFT is freely generated (see \cite{Argyres:2017tmj} for a discussion of the low energy consistency of this assumption).  In \cite{Argyres:2017tmj} it was also shown that CBs of SCFTs with non-freely generated chiral rings can have intricate complex singularities which can be separating and non-equi-dimensional --- thus making $\cC$ not even topologically a manifold --- but are not disallowed by any physical requirements.  Thus while it is conjectured that all $\cN=2$ SCFT CB chiral rings are freely generated, we do not know of a physical reason for this to be true. 

Even with the assumption that $\cC$ is a complex manifold, there are only a limited number of general things that can be physically inferred about the topology and analytic geometry of the set of metric singularities $\cV\subset\cC$  on the CB.
If a state in the theory with charge $\bq\neq0$ becomes massless at a point where $Z_\bq=0$, then there will be charged massless states in the spectrum of the effective theory everywhere on the locus $\cV_\bq := \{\bu\in\cC\, |\, Z_\bq(\bu)=0\}$.  This follows since if there were a wall of marginal stability transverse to the $Z_\bq=0$ locus for the BPS state with charge $\bq$ to decay, say, to states with charges $\bp$ and $\bm$, then charge conservation and marginal stability imply $Z_\bq = Z_\bp+Z_\bm$ and arg$(Z_\bq)=$ arg$(Z_\bp)=$ arg$(Z_\bm)$.  Therefore $Z_\bq=0$ implies $Z_\bp=Z_\bm=0$.

The set of all metric singularities $\cV$ will be the union\footnote{If $\cV_\bq$ itself has disconnected components, then it may be possible that only some of these components are in $\cV$, since then walls of marginal stability may prevent BPS states with charge $\bq$ from being in the spectrum of the effective theory at other components.} of the $\cV_\bq$ subsets, $\cV = \bigcup_{\bq\in \Phi}\cV_\bq$, for $\bq$ running over some subset, $\Phi$, of charges in the EM charge lattice $\L$.  Since the equation defining $\cV_\bq$ is linear in $\bq$, all $\bq\in \Phi$ can be taken to be primitive vectors in $\L$.  However $\Phi$ need not be a sublattice of $\L$, since if there are BPS states with charges $\bp$ and $\bq$ in the spectrum, there need not be a BPS state with charge $\bp+\bq$ in the spectrum, as the states with charges $\bp$ and $\bq$ in the spectrum need not be mutually BPS.

Since the section, $\s$, is a locally holomorphic function on $\cM$, so is $Z_\bq = \bq^T \s$, and therefore $\cV_\bq$ is a complex codimension one locus in $\cC$.  However, because $Z_\bq$ is not analytic on $\cC$ (it has branch points along $\cV_\bq$, reflecting its multivaluedness associated with its having non-trivial EM duality monodromy around $\cV_\bq$), $\cV_\bq$ is not obviously an analytic subspace of $\cC$.  In particular, a given $\cV_\bq$ might have accumulation points where it becomes dense in $\cC$, and, if the cardinality of $\Phi$ is infinite, then the union of the $\cV_\bq$ could conceivably also accumulate densely in $\cC$.
For example, if $u$ is one of the $r$ complex coordinates on $\cC$, one could imagine a central charge which behaves like $Z_\bq = \sqrt{\sin(\pi/u)}$.  This has zeros (and branch points) at the hyperplanes $u = 1/n$ for $n$ integer, and is dense around the $u=0$ hyperplane.  If a state of charge $\bq$ were in the spectrum, then $\cV$ would include all these hyperplanes.  Furthermore, by including the $u=0$ hyperplane in $\cV$ (for instance if  there were another state of charge $\bp$ in the spectrum with central charge, say, $Z_\bp = u^{1/3}$), then every point in $\cM = \cC\setminus\cV$ has an open neighborhood with $|Z_\bq|>0$ and $|Z_\bp|>0$, and so has a consistent low energy interpretation as a theory of free massless vector multiplets.

Of course the above toy example is not a full-fledged SK geometry at its regular points.  In particular, we suspect that there is no set of EM duality monodromies and compatible SK metric on $\cM$ consistent with $Z_\bq$ having an essential singularity at $u=0$.  Since we do not know how to prove it, we will assume that such behavior does not occur.  In particular, we will assume that 
\begin{align}\label{non-dense}
&\text{\emph{Any complex curve in $\cC$ transverse to $\cV$ intersects $\cV$ in}}\nonumber\\
&\text{\emph{a set of points with no accumulation point.}}
\end{align}
If $\cV$ were an analytic subset of $\cC$, this would essentially be the definition of it being of complex co-dimension $\ge 1$ in $\cC$. 

We now add superconformal invariance to the mix, thereby greatly constraining the topology and geometry of $\cV$.

\subsection{Complex scaling action and orbits in rank-2}
\label{cplxscaleact}

For the remainder of the paper we focus on CBs of $\cN=2$ SCFTs.  In particular, we will therefore only need to characterize those $\cV$ which are invariant (as a set, not pointwise) under superconformal transformations.

Conformal invariance, together with $\cN=2$ supersymmetry, implies that there is a $\tCs$ action on the CB which arises as follows:  scale invariance implies a smooth $\R^+$ action on $\cC$, arising from the action under dilation $D$, with an isolated fixed point at the unique superconformal vacuum, $O\in\cC$.  $\cN=2$ superconformal invariance implies that, in addition, there exists a $\U(1)_R$ global symmetry.  On the Coulomb branch the vevs of chiral scalars spontaneously break both $D$ and $\U(1)_R$, and their respective $D$ and $\U(1)_R$ charges are proportional.  This means that the $\R^+$ $D$-action and the $\R$ $\U(1)_R$ action\footnote{Note that we do not require that the $\U(1)_R$ action is a circle action, but only an $\R=\til{S^1}$ action.  This is equivalent to not requiring that the scaling dimensions of the coordinates on $\cC$ be rational.  In the end, however, we will only find solutions in which the dimensions are all rational.} on $\cC$ combine to give a holomorphic $\tCs$ action on $\cC$, which we denote by $P \mapsto \l\circ P$ for $\l\in\tCs$ and $P\in\cC$.  Here $\tCs$ denotes the universal cover of $\C^*$, e.g., the Riemann surface of $y=\ln\l$.  We will call this $\tCs$ action on $\cC$ the \emph{complex scaling} action on the CB.  We normalize the $\tCs$ action so that quantities with mass dimension 1 scale homogeneously with weight one in $\l$.

Let us specialize now to the case of a 2 complex dimensional CB.  Take $\bu:=(u,v)$ to be a vector of complex coordinates on an open set around $O$.  Without loss of generality we will take $O = (0,0)$.  In a neighborhood of $O$ there exists a continuous complex scaling action on $\cC$ which fixes $O$.  The scaling action can then be linearized around $O$, and then exponentiated to get an action of the form
\begin{align}\label{r2C*act}
\l\,\circ:\  
\bu \mapsto \l^M \, \bu\, , 
\quad \l \in \tCs\, ,
\quad M \in \GL(2,\C) \, .
\end{align}
Up to a complex linear change of basis, $M$ can be taken in Jordan normal form.  If it has a non-trivial Jordan block, $M = \left(\bsmat \D & 1 \\ 0 & \D \esmat\right)$, then \eqref{r2C*act} corresponds physically to a scaling action on the two complex scalar operators around $O$ on the CB which is not reducible.  Such non-reducible representations of the conformal algebra were shown in \cite{Mack:1975je} to not occur in unitary CFTs.  Therefore $M$ in \eqref{r2C*act} is diagonalizable, $M = \left(\bsmat \D_u & 0 \\ 0 & \D_v \esmat\right)$, giving the $\tCs$ action
\begin{align}\label{scaling-ii}
\l\,\circ:\ 
\bpmat u \\v \epmat 
\mapsto 
\bpmat \l^{\D_u} u \\ \l^{\D_v} v \epmat, 
\quad \l \in \tCs\, .
\end{align} 
This corresponds physically to the existence, in the spectrum of the SCFT/IRFT theory at the vacuum $O$, of a basis of CB scalar operators for which the scaling action reduces to that of two primary fields with definite scaling dimensions equal to $\D_u$ and $\D_v$.  Conformal invariance demands that these scaling dimensions be real and positive, and, since we have assumed via \eqref{freeCBring} that the CB chiral ring is freely generated, unitarity implies that they are also both greater than or equal to 1 (see \cite{Argyres:2017tmj} for a discussion):
\begin{align}\label{unitarity}
\D_u\ge 1 \quad\text{and}\quad \D_v\ge 1\, .
\end{align}
The positivity of $\D_u$ and $\D_v$ implies that any neighborhood of $O$ can be analytically continued to all of $\C^2$ using the exponentiated action \eqref{scaling-ii}.  Thus, as a complex space, $\cC=\C^2$, and $(u,v)\in \C^2$ are complex coordinates vanishing at the superconformal vacuum and diagonalizing the scaling action.

\paragraph{Complex scaling orbits and singularities.}

Since dilatations and $\U(1)_R$ transformations are symmetries of the SCFT, the complex scaling action \eqref{scaling-ii} on the CB must fix $\cV$ as a set.  Thus $\cV$ will be unions of orbits $\cV_i$ of this $\tCs$ action, and we write $\cV:=\left\{ \bigcup_i \cV_i\, |\, \l\circ\cV_i \simeq \cV_i \right\}$.

There are three qualitatively different 1-dimensional orbits of this complex scaling action: (a) the orbit through the point $(u,v)=(1,0)$, (b) the orbit through the point $(u,v)=(0,1)$, and (c) the orbit through a point $(u,v)=(\w,1)$ for $\w\neq0$.
\begin{itemize}
\item Type (a) is the submanifold $\cV_\infty :=\{v=0 \ \&\ u\neq 0\} \simeq \C^*$ consisting of the $v=0$ plane minus the origin.
\item Type (b) is the submanifold $\cV_0 :=\{u=0 \ \&\ v\neq 0\} \simeq \C^*$ consisting of the $u=0$ plane minus the origin.
\item Type (c) orbits are the non-zero solutions to the equation $\cV_\w:=\{u=\w \, v^{\D_u/\D_v} \}$ for a given $\w\in\C^*$. \label{orbitc}
\end{itemize}
Thus we can denote all the possible complex scaling orbits by $\cV_\w$ by allowing $\w\in\P^1 \simeq \{0\}\cup\C^*\cup\{\infty\}$.  We will call orbits of types (a) or (b) ``unknotted" orbits, and orbits of type (c) ``knotted" orbits, for reasons which will become clear in Section \ref{sec2.3}.  

Now assume that a knotted orbit $\cV_\w$, $\w\in\C^*$, is a component of the set of singularities $\cV$.  If $\D_u$ and $\D_v$ are not commensurate, then $\cV_\w$ does not satisfy our second regularity condition \eqref{non-dense}.  For instance, the intersection of $\cV_\w$ with the curve $u=\w$ has an accumulation point unless $\D_u/\D_v \in \Q$, i.e., unless $\D_u$ and $\D_v$ are commensurate.  Furthermore, when $\D_u$ and $\D_v$ are commensurate then the general variety of singularities is $\cV = \{0\} \cup_{i\in I} \cV_{\w_i}$ for some index set $I$. A necessary condition for the $\w_i$ not to have an accumulation point in $\P^1$ is that $I$ must be a finite set; that is $|I|<\infty$.  

Actually, it is interesting to note that while the regularity assumption \eqref{non-dense} is needed to deduce that the number of components $\cV_{\w_i}$ is finite, it is not needed to deduce that $\D_u$ and $\D_v$ are commensurate, so long as there is a knotted component (i.e., one with $\w \in \C^*$).   The argument is as follows:  If $\D_u/\D_v \notin\Q$, then $\cV_\w$ with $\w\in\C^*$ is dense in a 3-real-dimensional submanifold of $\cC$.  This is easy to see, for instance, by foliating $\cC$ by 3-spheres related by dilatations.  The intersection of the 3-sphere with $\cV_\w$ fixes $|u|$ and $|v|$, and imposes the linear constraint $\th = (\D_u/\D_v) \f$ on the phases $e^{i\th}$ and $e^{i\f}$ of $u$ and $v$, respectively.  Thus $\cV_\w \cap S^3$ is this line wrapping the ``square" torus, $T^2 = \{ (\th,\f) \, |\, \th \sim \th+2\pi \ \text{and}\ \f \sim\f+2\pi \}$.  If the slope $\D_u/\D_v$ of this line is irrational, then the line does not close, and is dense everywhere in $T^2$.  $\cV_\w$ is thus dense in the 3-manifold, $\cT_3$, which is the orbit of this $T^2$ under dilatations (this bit of analytic geometry will also be used in Section \ref{sec2.3}, where it is explained in more detail.)  Now pick any point $P\in\cT_3$ which is not on $\cV_\w$.  Then, because $\cV_\w$ is dense in $\cT_3$, every open neighborhood of $P$ intersects $\cV_\w$.  Thus there is no open neighborhood of $P$ with central charges bounded away from zero, and so $P$ cannot be consistently interpreted as a regular point on the CB --- i.e., as having a low energy description as a theory of free massless vector multiplets.  Thus $\cV_\w$ cannot be a component of $\cV$ for incommensurate $\D_u$ and $\D_v$.\footnote{There is a way to avoid this conclusion:  all points of $\cT_3$ could be in $\cV$.  This can happen if the uncountably infinite number of orbits $\cV_\w$ consisting of all $\w$ with fixed norm $|\w|$ are part of $\cV$.  This would violate the regularity assumption \eqref{non-dense}.}  This should be contrasted with the example given in the paragraph above \eqref{non-dense}. 

We have therefore learned that if $\D_u$ and $\D_v$ are commensurate, then the singularity set can be any union of the point at the origin with a finite number of distinct $\tCs$ orbits $\cV_\w$ (knotted or not), while if $\D_u$ and $\D_v$ are incommensurate, the singularity set can only be a union of the origin with either or both unknotted orbits ($\cV_0$ and $\cV_\infty$).

We will see eventually, in Section \ref{sec4.2}, that in the case where only unknotted orbits are present in $\cV$, the CB geometry factorizes into that of two decoupled rank-1 SCFTs.  Since the scaling dimensions of the CB parameters of rank-1 SCFTs are already known to be rational, we will thereby learn that in all cases $\D_u$ and $\D_v$ are commensurate.  So from now on we will write 
\begin{align}\label{Duvtopq}
\frac{p}{q} := \frac{\D_v}{\D_u}  
\qquad \text{for}\qquad 
p,q \in \Z^+
\qquad\text{with}\qquad 
\gcd(p,q) = 1\, .
\end{align}
$\cV_\w \cup \{0\}$ is thus the algebraic variety described by the equation
\begin{align}\label{Vwcurve}
u^p = \w v^q \, ,
\end{align}
and, algebraically, $\cV$ is described by the curve in $\cC=\C^2$:
\begin{align}\label{singularity}
\cV = \left\{ u^{\ell_0}\cdot 
\prod_{j=1}^\ell (u^p-\w_j v^q)
\cdot v^{\ell_\infty} = 0 \right\} ,
\end{align}
where the $\w_j\in\C^*$ are all distinct.  Here $\ell_0$ and $\ell_\infty$ are either 0 or 1, depending on which unknotted orbits are present, and $\ell$ is the number of knotted orbits in $\cV$.  In particular, $\cV\setminus\{0\}$ is a smoothly embedded 1-dimensional complex submanifold of $\cC$ with $\ell_0+\ell+\ell_\infty$ disconnected components.

\subsection{Topology of $\cV\subset\cC$}
\label{sec2.3}

We now describe the (point set) topology of how $\cV$ is embedded in $\cC$.  A given knotted component, $\cV_{\w_j}$ with $\w_j\in\C^*$, is homeomorphic to the curve $X(p,q) := \{u^p = v^q\} \subset \C^2$ simply by continuously mapping $\w_j$ to 1 in $\C^*$.  Likewise, the set of $\ell$ such distinct components is homeomorphic to the curve $X(p,q)^\ell := \{u^{p\ell} = v^{q\ell}\} \subset \C^2$ simply by continuously mapping each $\w_j$ to $e^{2\pi i j/\ell}$ along paths which do not intersect in $\C^*$.

To see the topology of $X(p,q)$, intersect it with $S^3_\r :=\{|u|^p + |v|^q = 2 e^\r\}$ for $\r\in\R$, which are a family of topological 3-spheres foliating $\C^2\setminus\{0\}$.  Note that different $\r$'s are related by dilations (i.e., $\l \in \C^* \cap \, \R^+$).  We then see that $X(p,q) \cap S^3_{\r=0}$ is a ``deformation retract'' of $X(p,q) \setminus \{ 0\}$ in $\C^2$.  Therefore $\pi_1 (\C^2\setminus X(p,q) ) \simeq \pi_1 (S^3_0 \setminus (X(p,q) \cap S^3_0) )$.  Therefore it is enough to analyze the topology of $X(p,q) \cap S^3_0$ in $S^3_0$.  Henceforth we will denote $X(p,q) \cap S^3_0 := K(p,q)$.  $K(p,q)$ is a one real-dimensional curve given by
\begin{align}\label{torusknot}
K(p,q) = \left\{(u,v)\in\C^2 \ |\  u = e^{i\th}, \quad v = e^{i\f} \quad\text{with}\quad p\th = q\f \mod 2\pi \right\}.
\end{align} 
Thus $K(p,q)$ is a knot in $S^3_0$ which lies on the 2-torus $T^2 := \{(u,v)\in\C^2\ |\ u = e^{i\th}, \ v = e^{i\f}\ \text{for}\ \th,\f\ \in \mathbb{R}\}$, embedded in $S^3_0$, and winds $p$ times around one cycle (the $\f$ or $v$ direction) and $q$ times around the other cycle (the $\th$ or $u$ direction).

A similar construction shows that $\pi_1 (\C^2\setminus X(p,q)^\ell ) \simeq \pi_1 (S^3_0 \setminus K(p,q)^\ell )$, where $K(p,q)^\ell$ is the link with $\ell$ components, each of which is homeomorphic to the $K(p,q)$ torus knot, but the $j$th component is translated along the $\th$ direction by $2\pi j/(p\ell)$.  Thus
\begin{align}\label{toruslink}
K(p,q)^\ell = \left\{(u,v)\in\C^2 \ |\  u = e^{i\th}, \quad v = e^{i\f} \quad\text{with}\quad p\th = q\f \mod 2\pi/\ell \right\}.
\end{align}
Finally, the intersections $K_0:= \cV_0\cap S^3_0$ and $K_\infty :=\cV_\infty\cap S^3_0$ are the circles (or ``unknots")
\begin{align}\label{unknots}
K_0 &= \left\{(u,v)\in\C^2 \ |\  
u = 0, \qquad\quad \,
v = 2^{1/q} e^{i\f} \quad
\text{with}\quad \f\in\R \mod 2\pi\right\},\nonumber\\
K_\infty &= \left\{(u,v)\in\C^2 \ |\  
u = 2^{1/p} e^{i\th}, \quad 
v = 0 \qquad\quad\,
\text{with}\quad \th\in\R \mod 2\pi\right\}.
\end{align}

We denote the total link consisting of a torus link together with  unknots by
\begin{align}\label{genlink}
L_{(p,q)}(\ell_0,\ell,\ell_\infty) := (K_0)^{\ell_0} \cup K(p,q)^\ell \cup (K_\infty)^{\ell_\infty}.
\end{align}
Here we are using a notation where $(K_0)^{\ell_0} := K_0$ if $\ell_0= 1$, and $:=\varnothing$ if $\ell_0=0$, and similarly for $(K_\infty)^{\ell_\infty}$.  Similarly, $\ell=0$ means that there is no torus link component.  Thus, for example, $L_{(p,q)}(0,\ell,0) = K(p,q)^\ell$, and $L_{(p,q)}(0,0,1) = K_\infty$.  

\begin{figure}[ht]
\centering
\includegraphics[width=.50\textwidth]{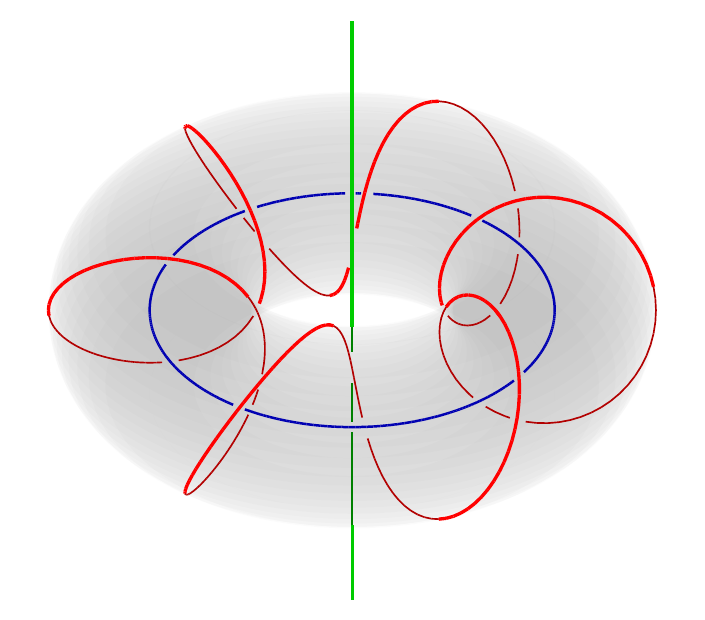}
\caption{Depiction of an $L_{(1,6)}(1,1,1)$ link consisting of the blue ($K_0$), red ($K(1,6)$), and green ($K_\infty$) circles.  The solid gray torus is there for visualization purposes.}
\label{knot1}
\end{figure}

These links are relatively easy to visualize.  For example, Figure \ref{knot1} depicts an $L_{(1,6)}(1, 1, 1)$ link with the $K(1,6)$ knot in red on the surface of a solid gray torus (the torus is present purely for visualization), the $K_0$ threading the interior of the torus in blue, and $K_\infty$ as the ``z-axis" in green.  The three dimensions are the stereographic projection of $S^3_0$ to $\R^3$ with the point at infinity being $(u,v)=(-2^{1/p},0)$ and origin being $(u,v)=(+2^{1/p},0)$.   Thus the green line goes through the point at infinity, so is topologically a circle.

\paragraph{The fundamental group of $\cC\setminus\cV$.}

The fundamental group of the metrically smooth part of the CB $\cM$, with $\cV$ given in \eqref{singularity} is $\pi_1(\cM) = \pi_1(S_0^3 \setminus L_{(p,q)}(\ell_0,\ell,\ell_\infty))$.  The last expression is known as the \emph{knot group} of the  link \eqref{genlink}.  

\begin{figure}[ht]
\centering
\includegraphics[width=.50\textwidth]{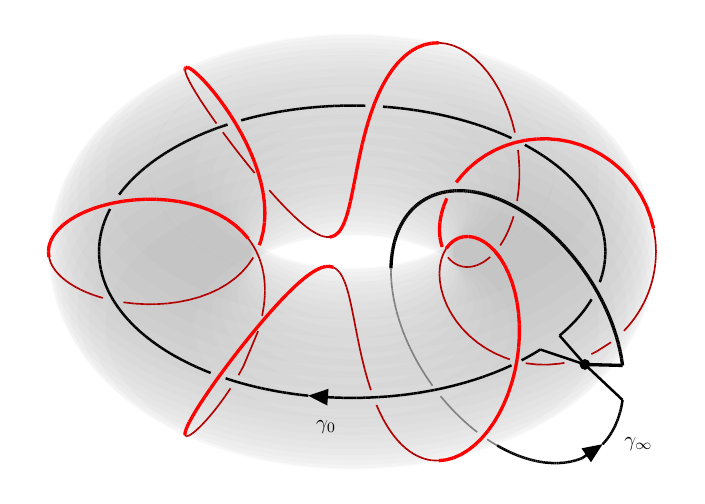}
\caption{Depiction of an $L_{(1,6)}(0,1,0)$ link consisting of  the red circle.  The $\g_0$ cycle threads the interior of the donut, while $\g_\infty$ threads the hole of the donut.}
\label{knot2}
\end{figure}

One can compute the knot group using the groupoid Seifert-van Kampen theorem \cite{Argyres:2019kpy}.  For clarity, we first describe the result in the case with a single torus knot and no unknots.  It is
\begin{align}\label{L(010)}
\pi_1(\cM) = \langle\ \g_0,\g_\infty\ |\ 
{\g_0}^p= {\g_\infty}^q\ \rangle\, .
\end{align}
Here the fundamental group has been given as a set of generators, $\g_0$ and $\g_\infty$, subject to a single relation, $\g_0^p = \g_\infty^q$.  This is the classic result for a torus knot found from a simple application of the Seifert-van Kampen theorem \cite{Hatcher:2002}.  The $\g_0$ and $\g_\infty$ cycles are shown in the example of a $K(1,6)$ knot in Figure \ref{knot2}.  The relation, $\g_0 = {\g_\infty}^6$, is obvious in this simple case. 

\begin{figure}[ht]
\centering
\includegraphics[width=.50\textwidth]{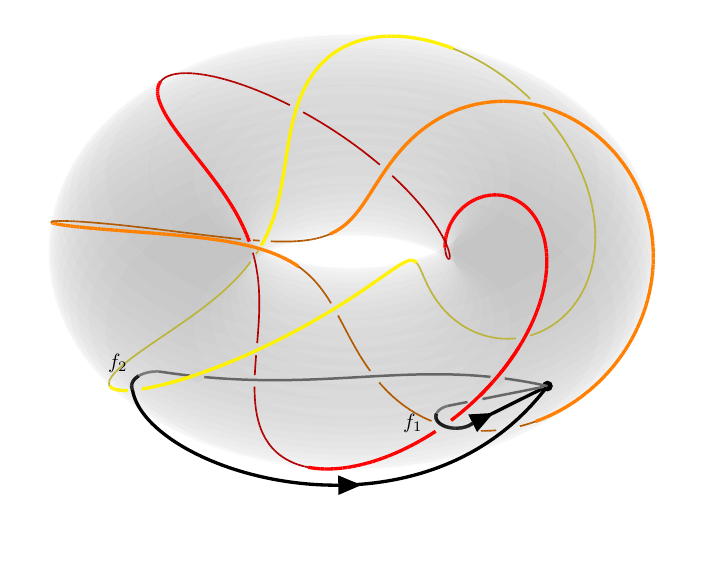}
\caption{Depiction of an $L_{(1,2)}(0,3,0)$ link consisting of the red, orange, and yellow circles.  The $f_1$ cycle links the first strand in the direction of the $\g_0$ cycle, while $f_2$ links the first two strands.  The $\g_0$ and $\g_\infty$ cycles, as in Figure \ref{knot2}, are not shown.}
\label{knot3}
\end{figure}

The generalization to the case of a torus link, $K(p,q)^\ell$, is quite non-trivial, but thanks to the analysis in \cite{Argyres:2019kpy} we have the following result:
\begin{align}\label{L(0l0)}
\pi_1(\cM) = \langle\ \g_0,f_1,f_2,\ldots,f_\ell,\g_\infty\ |\ 
{\g_0}^p f_j= f_j {\g_\infty}^q \, ,\, f_\ell=1\ \rangle\, .
\end{align}
There are $\ell-1$ additional generators, $f_j$ for $j=1,\ldots,\ell-1$, and $\ell$ relations.  It is convenient to add an $\ell$th additional generator, $f_\ell$, simply to make the set of relations look more uniform, but then we must impose $f_\ell=1$.
The $f_j$ generators correspond to cycles which loop individual strands of the link, as shown in Figure \ref{knot3} for the case of a $K(1,2)^3$ link.

\begin{figure}[ht]
\centering
\includegraphics[width=.50\textwidth]{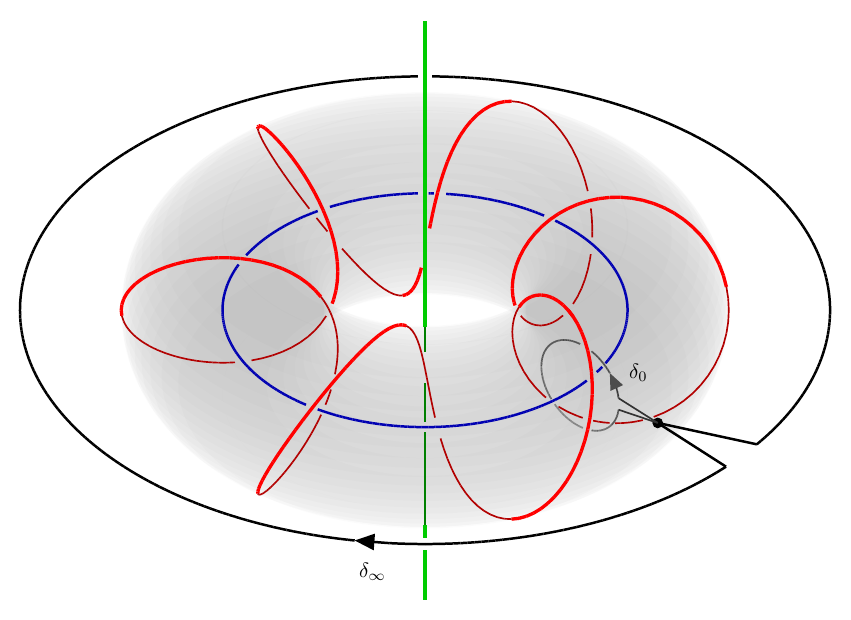}
\caption{Depiction of an $L_{(1,6)}(1,1,1)$ link consisting of the blue ($K_0$), red ($K(1,6)$), and green ($K_\infty$) circles.   The $\d_0$ cycle links only the $K_0$ unknot, while $\d_\infty$ links only the $K_\infty$ unknot.  The $\g_0$ and $\g_\infty$ cycles, as in Figure \ref{knot2}, are not shown.}
\label{knot4}
\end{figure}

In \cite{Argyres:2019kpy} the general result with unknots was found to be:
\begin{align}\label{L(1K1)}
\pi_1(\cM) &= \langle\ \d_0,\g_0,f_1,\ldots,f_\ell,\g_\infty,\d_\infty\ | \ 
\g_0\d_0=\d_0\g_0 \, , \, 
\g_\infty\d_\infty=\d_\infty\g_\infty \, ,
\nonumber\\
& \qquad \qquad \qquad
{\g_0}^p {\d_0}^q f_j= f_j {\d_\infty}^p{\g_\infty}^q\, ,\, 
f_\ell=1\ \rangle\, .
\end{align}
The two $\d$ generating cycles associated with the unknots are depicted in Figure \ref{knot4}.  Note that if $\ell_0$ or $\ell_\infty$ (or both) are zero, indicating the absence of one or both of the unknot singularities, then the general result \eqref{L(1K1)} holds but with additional relations setting $\d_0$ or $\d_\infty$ (or both) equal to the identity.

A set of consistent EM duality monodromies around the components of $\cV$ must form a representation of $\pi_1(\cM)$ in $\SpDtZ$ (the EM duality group).  The EM duality monodromy around a given component of $\cV$ largely determines the analytic form of the section $\s$ of special coordinates on the CB near $\cV$;  we will explain this in Section \ref{secSKmetric} below. A representation of $\pi_1(\cM)$ in $\SpDtZ$ is then arithmetic ``data" constraining the possible global form of the CB geometry:  it provides the boundary conditions that an analytic continuation of $\s$ from the vicinity of one component of $\cV$ to that of another must satisfy.  The rest of this paper is aimed at sorting out the ingredients necessary for performing this analytic continuation.

\section{A few concrete examples}\label{phys}

Since the discussion in the previous section might appear quite abstract, we will now illustrate the singularity structure of a few CBs with some familiar (i.e., lagrangian) rank-2 SCFTs.  This will provide a direct physical interpretation of the topology of $\cV \subset \cC$.  In particular, we will analyze the singularity structure of two well-known rank-2 theories: $\SU(3)$ gauge theory with a single massless hypermultiplet in the adjoint representation, and $\SU(3)$ gauge theory with six massless  hypermultiplets in the fundamental representation.  These examples are particularly illuminating, given that the singularity structure of these two theories realize all the possible distinct topologies discussed above, namely unknots, single $(p,q)$ knots, and a $(p,q)$ link.

The moduli space of a lagrangian theory can be explicitly constructed from its field content.  $\cN=2$ gauge theories are described in terms of $\cN=1$ superfields by a chiral field strength multiplet $W=W^a T^a$, and a chiral multiplet $\F = \F^a T^a$, both transforming in the adjoint representation of the gauge group, which form an $\cN=2$ vector multiplet, and chiral multiplets $Q^i_I$ and $\tQ_i^I$ in representations of the gauge group $\bR_Q$ and $\bar\bR_Q$, which form a hypermultiplet.  The index $a$ runs over a gauge Lie algebra basis, $I=1,\ldots,\dim \bR_Q$ is the hypermultiplet gauge representation index, and $i$ is a flavor index; $i$ distinguishes different hypermultiplets in the same representation $\bR_Q$. 

We begin by describing some generalities about $\SU(3)$ CBs.  The CB is parametrized by the vacuum expectation values of $A$, the complex scalar in $\F$.  To simplify notation we use the symbol $A$ in place of $\vev A$ where it will not be confusing.  Upon eliminating the auxiliary fields, the $\cN=2$ lagrangian contains a scalar potential $V \sim \Tr \big([A,A^\dag]\big)^2$, which implies that the Coulomb vacua are parametrized by $A$ taking value in the complexified Cartan subalgebra, and so can all be simultaneously diagonalized by a gauge rotation.  In particular, for $\SU(3)$ we can write:
\beq\label{SU3CB}
A= \bpmat a_1&&\\ &a_2&\\ &&a_3 \epmat ,
\qquad \sum_{k=1}^3 a_k=0 \, .
\eeq
The $a_k$'s are not gauge invariant, and the residual gauge action on \eqref{SU3CB} corresponds to the Weyl group of the gauge Lie algebra, which is just the group of permutations of the $a_k$.   The gauge-invariant coordinates on $\cC$ are the algebraically independent Weyl invariant combinations of the $a_k$'s,
\begin{align}\label{}
u := \frac16 \sum_k a_k^2\, ,
\qquad
v:= \frac12 a_1a_2a_3\, ,
\end{align}
where the overall normalization of $u$ and $v$ is arbitrary, and has been chosen to simplify the expressions below.

We can fix the Weyl group redundancy in the description \eqref{SU3CB} by restricting the $a_k$'s to a single Weyl chamber by setting  $A \cdot \bal_{1,2} \geq 0$, where $\bal_{1,2}$ are the $\SU(3)$ simple roots.  In the matrix notation of \eqref{SU3CB}, the simple roots can be represented by
\begin{align}\label{}
\bal_1 := \left(\bsmat 1&&\\&-1&\\&&0\esmat\right),
\qquad
\bal_2 := \left(\bsmat 0&&\\&1&\\&&-1\esmat\right),
\end{align}
and the dot product is the matrix trace.  Then the Weyl chamber conditions correspond to setting $a_1\geq a_2\geq a_3$.  

The CB vev \eqref{SU3CB} generically breaks the gauge group to $\U(1)^2$ unless two of the $a_k$'s coincide, in which case one of the two $\U(1)$'s is enhanced to an $\SU(2)$.  This happens precisely at the boundary of the Weyl chamber which is given by those values of $A$ for which $A\cdot \bal_{1,2}=0$.

The theory also contains $\cN=1$ superpotential terms $\cW \sim \tQ_i \F Q^i$, where the $T^a$'s act in the appropriate representation on $(\tQ_i,Q^i)$.  When $A$ acquires a vev, the superpotential generates masses for the hypermultiplets; in particular, for the fermionic components (which to make notation easier we will also indicate with $\tQ_i$ and $Q^i$) the mass term is of the form
\beq
m_I \sim \tQ_i^I \big(A\cdot \bmu_I\big) Q^i_I\, .
\eeq
The $\bmu_I$ run over the weight vectors of the representation $\bR_Q$.  Thus on an interior point of the Weyl chamber, unless $A \cdot \bmu_I = 0$ for some $I$, all hypermultiplets are massive, and the effective theory on the CB is a free $\U(1)^2$ theory. 

As stated previously, the singular locus $\cV\in\cC$ is parametrized by those $(u,v)$ for which extra massless states charged under the $\U(1)$'s appear in the theory.  From the discussion above we see this happens for those values of $A$ such that
\begin{itemize}
\item[$(a)$] $A\cdot\bmu_I = 0$: some component of the hypermultiplets become massless, or
\item[$(b)$] $A\cdot\bal_{1,2} = 0$: $W^\pm$ bosons associated with the extra unbroken $\SU(2)$ become massless, restoring an $\SU(2)$ gauge symmetry.
\end{itemize}

\subsection{$\SU(3)$ with 1 adjoint hypermultiplet}

In this example the theory contains only one hypermultiplet, transforming in the adjoint representation of $\SU(3)$.  In fact, this theory has an enhanced $\cN=4$ supersymmetry.  The weight vectors of the representation of the hypermultiplet obviously coincide with the roots of the Lie algebra,
and therefore along the (singular) subvariety where one of the two $\U(1)$'s is enhanced to a non-abelian $\SU(2)$, some components of the hypermultiplet also become massless.  Before analyzing the effective IR theory along this subvariety, we write it explicitly in terms of the coordinates $(u,v)$ on $\cC$:
\beq\label{extraSU2}
\left.\begin{array}{c}
A^1_{\SU(2)}\cdot\a_1 = 0
\quad {\rm or} \quad 
A^1_{\SU(2)}=\left(\bsmat a&&\\ &a&\\ &&-2a\esmat\right)
\\[2mm]
A^2_{\SU(2)}\cdot\a_2 = 0
\quad {\rm or} \quad 
A^2_{\SU(2)}=\left(\bsmat 2a&&\\ &-a&\\ &&-a\esmat\right)
\end{array} 
\right\}
\qquad \Longrightarrow \qquad u^3=v^2\, .
\eeq
In the notation introduced in Section \ref{cplxscaleact}, the hypersurface $u^3=v^2$ (minus the origin) is a knotted $\tCs$ orbit of type $(c)$, and it is topologically equivalent to $K(2,3)$.

The components of the hypermultiplets which are massless along \eqref{extraSU2} transform in the adjoint representation of the unbroken $\SU(2)$, and are uncharged under the other $\U(1)$ factor.   It follows that the effective theory along \eqref{extraSU2} is an $\cN=4$ $\SU(2)$ gauge theory with a decoupled free $\U(1)$ factor.  The existence of a SCFT all along \eqref{extraSU2} implies the presence of metric singularities all along the hypersurface.  It follows that in this case $\cV$ is topologically equivalent to $L_{(2,3)}(0,1,0)$.

\subsection{$\SU(3)$ with 6 fundamental hypermultiplets}

This case is slightly more subtle.  The hypermultiplets transform in the fundamental representation of $\SU(3)$ whose weights are
\begin{align}\label{}
\bmu_1 = \frac13 \left(\bsmat 2&&\\&-1&\\&&-1\esmat\right),
\quad
\bmu_2 = \frac13 \left(\bsmat -1&&\\&2&\\&&-1\esmat\right),
\quad
\bmu_3 = \frac13 \left(\bsmat -1&&\\&-1&\\&&2\esmat\right).
\end{align}
Thus $A \cdot \bmu_I = a_I$, $I=1,2,3$, and therefore components of the hypermultiplets become massless if any of the $a_k$'s vanish.  Note that since we are working in a specific Weyl chamber, the only possibility for an $a_k$ to vanish away from the SCFT vacuum at the origin is:
\beq\label{unknotSU3}
A_0 \cdot \bmu_2 = 0
\quad {\rm or} \quad 
A_0=\bpmat a&&\\ &0&\\ &&-a \epmat
\qquad \Longrightarrow \qquad v=0\, .
\eeq
The hyperplane above is again one of the $\tCs$ orbits previously analyzed, specifically a type $(a)$ unknotted orbit.  

It is straightforward to analyze the effective IR description of the theory along \eqref{unknotSU3}.  The gauge group is fully broken to $\U(1)^2$, which can be chosen in such a way that the extra massless components are 6 massless hypermultiplets with charge 1.  In this case this is an IR free theory with massless matter all along \eqref{unknotSU3}, and we thus expect metric singularities along this sublocus.  Thus this provides a component of the singular locus, $\cV_0$, which is topologically equivalent to the link $L_{(2,3)}(0,0,1)$.

Note that this topological description misses the algebraic multiplicity of the singularity which can instead be inferred from the SW curve of the theory \cite{Argyres:1995wt, Landsteiner:1998pb, Argyres:2005pp}, where it is found to be of multiplicity 6.  This extra piece of information reflects the fact that 6 charge-1 hypermultiplets are becoming massless there, so the coefficient of the beta function of the $\U(1)$ gauge factor they are charged under is 6.

Now let us focus on those regions with an enhanced $\SU(2)$ symmetry and the corresponding effective theory.  It can be explicitly seen from \eqref{extraSU2} that away from the origin, none of the $a$'s vanish along this subvariety, thus below the energy scale $a$ all the hypermultiplets are massive.  The IR theory is a product of a pure $\SU(2)$ gauge theory with a decoupled free $\U(1)$.  Because the pure $\SU(2)$ is an asymptotically free theory, determining the location of the singular subvariety is trickier.  It is in fact well-known that the $\SU(2)$ confines at some scale $\L_{\SU(2)}$, and no massless W-bosons arise in the IR.  However, this theory still has a non trivial singularity structure; by appropriately tuning the CB vev of the pure $\SU(2)$ gauge theory, either a dyon or a monopole can become massless.  This is the celebrated result \cite{sw1} that the pure $\SU(2)$ theory has singularities at $\til a^2=\pm \L_{\SU(2)}^2$, where $\pm\til a$ are the vevs of the vector multiplet complex scalar in the $\SU(2)$ Cartan subalgebra.  Let us now turn to the implications of this observation for the singularity structure of the $\SU(3)$ theory.

We first need to relate $\til a$, parametrizing the IR $\SU(2)$ CB vev, with the $a_k$'s in \eqref{SU3CB}.  Notice that for $A\cdot\bal_1=0$ ($A\cdot\bal_2=0$) the IR $\SU(2)$ is embedded in the top left (bottom right) $2\times2$ corner of the $\SU(3)$ matrices.  Thus by inspection $\til a= (a_1-a_2)/2$ ($\til a= (a_2-a_3)/2$).  Next, observe that $\L_{\SU(2)}$, the confining scale for the pure $\SU(2)$ gauge factor, is proportional to the value of $a$ in \eqref{SU3CB}.  This can be seen as follows.  The strong coupling scale for an asymptotically free theory is defined as $\L \propto \mu \exp\{2\pi i\t_{\SU(2)}(\m)\}$, where $\m$ is an arbitrary scale at which the running gauge coupling of the $\SU(2)$ effective gauge factor has value $\t_{\SU(2)}(\m)$.  In the UV, the $\SU(3)$ theory is a SCFT, and so its gauge coupling, $\t$, is an exactly marginal coupling which therefore does not run with scaling.  Therefore at the scale $a$ where the $\SU(3)$ is Higgsed to $\SU(2)\times \U(1)$, the $\SU(2)$ effective coupling is $\t$:  $\t_{\SU(2)}(a) = \t$.  Therefore $\L_{\SU(2)}\propto a\, {\rm e}^{2\pi i\t}$. 

Now let us go back to the study of the singular variety of the $\cN=2$ $\SU(3)$ SCFT.  Confinement of the $\SU(2)$ implies that the region in \eqref{extraSU2} is no longer singular as there are no extra massless BPS states there.  Instead we expect a massless dyon and a massless monopole to enter the theory at $\til a^2 = \pm \L_{\SU(2)}^2$ which translates to the loci of adjoint scalar vevs:
\beq\label{singSU3}
\begin{array}{c}
A^1_\L=\bpmat a(1+\e)&&\\ &a(1-\e)&\\ &&-2a \epmat
\quad {\rm and} \quad
A^1_{i\L}=\bpmat a(1+i\e)&&\\ &a(1-i\e)&\\ &&-2a \epmat\\
{\rm or}\\
A^2_\L=\bpmat 2a&&\\ &-a(1+\e)&\\ &&-a(1-\e)\epmat 
\quad {\rm and}\quad
A^2_{i\L}=\bpmat 2a&&\\ &-a(1+i\e)&\\ &&-a(1-i\e) \epmat 
\end{array}
\eeq
where $\e={\rm e}^{2\pi i\t}$.  The singular subvarieties above can be parametrized in terms of $(u,v)$ coordinates as follows:
\beq
A^{1,2}_\L := \left\{u^3=\frac{1+\e^3/3}{1-\e^2/2}v^2\right\}, \qquad 
A^{1,2}_{i\L} := \left\{u^3=\frac{1-\e^3/3}{1+\e^2/2}v^2\right\} .
\eeq 
We call the union of these two components of the singular region $\cV_{\SU(2)}$, and it is topologically equivalent to 2 parallel $K(2,3)$ knots or an $L_{(2,3)}(0,2,0)$ link.

Thus the singular CB locus of the $\SU(3)$ with six fundamentals SCFT is the union of the $\tCs$ orbits described above: $\cV = \cV_0 \cup \cV_{\SU(2)}$.  It is topologically equivalent to an $L_{(2,3)}(0,2,1)$ link.  This result agrees with the more straightforward analysis of \cite{Argyres:1995wt, Landsteiner:1998pb, Argyres:2005pp} in which the SW curve for this theory is constructed and the discriminant locus computed explicitly.

\subsection{Other rank-2 lagrangian SCFTs}

A similar analysis can be performed for the other lagrangian rank-2 SCFTs.  There are quite a few possibilities.  In fact for each one of the semisimple rank-2 gauge algebras --- $\SU(3)$, $\SO(5)=\Sp(4)$, $\SU(2)\times \SU(2)$, and $G_2$ --- there are many allowed choices for hypermultiplet representations giving vanishing beta function for the gauge coupling.

The analysis of the singular geometries for all of these theories contains ingredients similar to the discussion just outlined above, and thus we will not present it in detail.  Still it is worth pointing out a few distinct features which we learn from the study of the CB geometries of lagrangian SCFTs:

\begin{itemize}

\item The singular locus $\cV^{\cN=4}_\gf$ of the CB geometries for theories with enhanced $\cN=4$ supersymmetry and gauge Lie algebra $\gf$ are topologically $L_{(2,n)}(0,1,0)$ links, where $n$ is the highest dimension of the Casimir of Weyl$(\gf)$.  Furthermore the CB in this case is an orbifold $\cC^{\cN=4}_\gf = \C^2/\G$, where $\G={\rm Weyl}(\gf)$, and $\cV^{\cN=4}_\gf$ corresponds to the fix points of the $\G$ action.  This is not the case for theories with only $\cN=2$ supersymmetry. 

\item In rank-2, as implied by the previous observation, the scale invariant limit of the CB geometry is sensitive to supersymmetry enhancement.  The singularity structure of theories with the same gauge group but enhanced $\cN=4$ are distinct from the ones with only $\cN=2$.  In rank-1 this was known not to be the case since the beginning \cite{sw1, sw2}.  

\item But, as in rank-1, many distinct rank-2 lagrangian SCFTs share the same scale invariant CB geometry.  For a given gauge group, there are multiple choices of hypermultiplet representation which give $\cN=2$ SCFTs.  In particular for $\SU(3)$, in addition to the two cases presented above, the theory with one hypermultiplet in the fundamental and one in a two-index symmetric representation is also a SCFT.  This theory has the same CB geometry as does the theory with six fundamentals.\footnote{We thank Y. L\"u for pointing this out to us.}  This is also the case for $\SO(5)=\Sp(4)$ gauge algebras where there are a few different representation assignments giving rise to $\cN=2$ SCFTs, all of which have singular loci topologically equivalent to $L_{(2,4)}(0,2,1)$, as is readily obtained from their SW curves \cite{Argyres:1995fw, Argyres:2005wx}.
\end{itemize}

The last point suggests that to fully distinguish the different SCFTs purely from the analysis of their CB geometries we need to study the allowed mass deformations of the scale invariant geometries.  This turned out to be a very fruitful effort in rank-1 \cite{paper1, paper2, allm1602, am1604, Argyres:2016ccharges}, but many of the techniques that worked there do not seem to generalize to rank-2.  We will not make any attempt to study mass deformations here but hope to study this problem in the future.

\section{SK geometry of the Coulomb branch in rank-2}
\label{secSKmetric}

In this section we will discuss constraints on the CB geometry that arise from demanding a regular special K\"ahler metric at all points of $\cM$.  In particular, after a brief review of the SK metric and integrability condition in Section \ref{sec4.0}, we will see in Section \ref{sec4.1} how the physical condition that the CB metric be regular in directions parallel to the singularity $\cV$ gives strong constraints on the possible EM duality monodromy around a path linking $\cV$.   

In Section \ref{sec4.2} we will use the results of Section \ref{sec4.1} to find the spectrum of possible dimensions $\{\D_u,\D_v\}$ of CB coordinates in the case where $\cV$ has no knotted components.  In particular, we show that the problem essentially factorizes into a product of rank-1 geometries, and so the allowed values of $\D_{u,v}$ are just those of the rank-1 CBs, recorded in Table \ref{table:eps} of appendix \ref{appRk1}.  These eight possible values are rational, and so $\D_u/\D_v$ are also rational.  This then completes the argument started in Section \ref{cplxscaleact} that the CB scaling dimensions are commensurate.

An important ingredient in the argument of Section \ref{sec4.2} is the use of monodromy around cycles which are orbits of the $\U(1)_R$ symmetry action on the CB.  Such monodromies necessarily have an eigenvalue with unit norm.  We call these ``$\U(1)_R$ monodromies" and explore them further in Section \ref{sec4.3}. We will indicate $\U(1)_R$ monodromies with a fancy $\mtM$. Since we have determined that the CB scaling dimensions are commensurate, there will be closed $\U(1)_R$ orbits through every point of the CB.  This, together with the SK integrability condition and regularity of the CB metric, implies that the eigenspace of the unit-norm eigenvalue of a $\U(1)_R$ monodromy $\mtM$ must contain a lagrangian subspace of the charge space, $V\simeq \C^4$.  This puts a strong constraint on the conjugacy class of $\mtM$.  In particular, using some results on the classification of $\Sp(4,\R)$ conjugacy classes reviewed in appendix \ref{appB0}, this shows that \emph{all} the eigenvalues of $M$ must have unit norm.  

\subsection{SK geometry of $\cM$}\label{sec4.0}

The condition that the K\"ahler metric be positive definite on the CB (required by unitarity of the effective theory on the CB) and the SK integrability condition can be translated into statements about the symplectic geometry of the subspaces of $V^*$ spanned by derivatives of the SK section $\s$.  

On $\cM$, the SK manifold of metrically regular points of $\cC$, the K\"ahler potential is given by \eqref{Kpot}, implying that the  K\"ahler form and hermitian metric on $\cM$ are written as
\begin{align}\label{Kform}
\w &= \vev{d\s \, \overset{\^}{,} \, d\sb}\, ,
\qquad
h = i \vev{d\sb\, \overset{\otimes}{,}\, d\s}\, ,
\end{align}
where $d$ is the exterior derivative on $\cM$ and $\vev{\,\cdot\, \overset{\^}{,} \,\cdot\,}$ means take the exterior product as forms on $\cM$ as well as evaluate the Dirac pairing on $V^*\simeq\C^4$ (the dual charge space in which the SK section takes its values).  Thus, in terms of good complex coordinates, $u^j$, $j=1,2$, in the neighborhood of any point of $\cM$, we have $\w=i h_{j\bar k} du^j \^ d\ub^k$ and $h = h_{j\bar k} du^j \otimes d\ub^k$ with:
\begin{align}\label{hmetric}
h_{j\bar k} := i \vev{\del_j\s,\delb_k\sb}\, ,
\end{align}
where $\del_i := \del/\del u^i$.

Positivity of the K\"ahler metric is equivalent to the conditions
\begin{align}\label{positivity}
h_{1\bar 1} >0\, , \quad
h_{2\bar 2} >0\, , \quad
\text{and}\quad
\det(h) >0\, .
\end{align}
In particular, the first two conditions imply from \eqref{hmetric} that
\begin{align}\label{pos1}
\vev{\del_j\s,\delb_j\sb} \neq 0
\quad\text{for}\ j=1,2\, .
\end{align}
Denote by $S_j$ the subspace of $V^*$ spanned by $\del_j\s$ and $\delb_j\sb$ at a given point of $\cM$.  Then \eqref{pos1} is equivalent to the statement that each $S_j$ is a 2-dimensional symplectic subspace\footnote{Recall that a dimension-$2s$ subspace, $S$, of a $2r$-dimensional symplectic vector space $V^*$ is symplectic if $\vev{\cdot,\cdot}$ restricts to a non-degenerate form on $S$.} of $V^*$.

The third condition in \eqref{positivity} implies that the top form on $\cM$ given by $\w^2$ does not vanish.  It follows from \eqref{hmetric} that $\w^2 \sim \e^{a_1 a_2 b_1 b_2}\, \delb_1 \sb_{a_1} \, \delb_2 \sb_{a_2} \, \del_1 \s_{b_1} \, \del_2 \s_{b_2} \, d^4u$.  The antisymmetrization on the $V^*$ indices comes from the fact that $J_\D^{-1} \^ J_\D^{-1} \propto \e$, where $J_\D^{-1}$ is the symplectic form on $V^*$ defined by the induced Dirac pairing.  Therefore $\w^2\neq0$ implies, in addition, only that $\del_j\s$, $\delb_j\s$, $j=1,2$, span all of $V^*$ at each point of $\cM$.  Thus, in particular, we learn that the dual charge space decomposes as
\begin{align}\label{V*decomp}
V^* = S_1 \oplus S_2\, .
\end{align}

The SK integrability condition \eqref{SKintegrability} is, in these coordinates, the statement that $\vev{\del_1\s , \del_2\s} =0$, i.e., that $\del_1\s$ and $\del_2\s$ span a lagrangian subspace\footnote{Recall that a dimension-$r$ subspace, $L$, of a $2r$-dimensional symplectic vector space $V^*$ is lagrangian if $\vev{\bv,\bw}=0$ for all $\bv, \bw\in L$.  ($r=2$ in this paper.)} of $V^*$, and therefore similarly for their complex conjugates.  This does not imply that $S_1$ and $S_2$ are symplectic complements\footnote{The symplectic complement of $S$ is defined by $S^\perp := \{ \bv\in V^* \, |\, \vev{\bv,\bw}=0\ \text{for all}\ \bw\in S\}$.} in $V^*$, but the integrability condition does imply that it is possible to pick special  $u^1$ and $u^2$ coordinates --- locally chosen to satisfy $\vev{\del_1\s,\delb_2\sb}=0$ --- for which $(S_1)^\perp = S_2$.

These simple relations tie together the symplectic geometry of the dual charge space, $V^*$, with the complex geometry of the metrically regular part $\cM$ of the CB.  We will now explore how and to what extent these relations extend to the metric singularities $\cV$ of the CB.

\subsection{SK geometry near $\cV$}\label{sec4.1}

A basic constraint on the SK geometry of a rank-$r$ CB at its singular locus $\cV$ is that the SK section $\s$ cannot diverge there.  For if (some components of) $\s$ did diverge at a point $P\in\cV$, then the set of charges $\bp$ such that $Z_\bp(P)<\infty$ would form a sublattice $\L'\subset\L$ of rank smaller than $2r$.  This would imply that all states with $\bq\in \L\setminus\L'$ decouple not just from the low energy theory, but from the theory as a whole (at all scales).  The decoupling of all states with charges in $\L'$ from the theory at arbitrarily high energy scales is a microscopic property of the theory.  Thus, by locality, it must be true of the theory at all its vacua.  The problem with this is the following:  the sublattice $\L'$ of allowed states will not be magnetically charged (in some duality frame) under at least one of the $\U(1)$ gauge factors.  This means that this $\U(1)$ gauge factor is either free and completely decoupled (if there are no states electrically or magnetically charged with respect to it) or UV incomplete (if there are some states electrically charged with respect to it).  We reject these behaviors because a completely decoupled free factor is uninteresting, and a UV incomplete factor will give rise to ``Landau poles" --- non-unitary behavior at high-enough scales.\footnote{Any power-law or even logarithmic divergence in $\s$ as one approaches a point in $\cV$ naively implies a  pole-or-stronger divergence in the K\"ahler line element, and thus an infinite metric distance to $\cV$.  This would be a contradiction since, by definition, the points of $\cV$ are at finite distance.  But this conclusion is naive because we can have $\vev{\bar\s_\infty,\s_\infty}=0$ for the divergent component, $\s_\infty$, of $\s$ without having $\s_\infty=0$ --- i.e., $\bar\s_\infty$ and $\s_\infty$ may be vectors in the same lagrangian subspace.  Thus requiring finiteness of $\s$ at $\cV$ (and thus everywhere on the CB) is a stronger condition than $\cV$ not being at metric infinity. 

Physical intuition leads us to expect that the $\s$-finiteness condition should be able to be derived from the other conditions in the sense that one can show that if there is a divergence, then either (a) it violates the not-at-metric-infinity requirement, or (b) it implies a violation of the positivity of the metric somewhere else on the CB, reflecting the Landau poles of the UV-incomplete theory.  But (b) is a non-local property of the CB geometry which we (the authors) do not have the tools to analyze.} 

Since $\s$ is holomorphic away from $\cV$ and does not diverge at $\cV$, it will have a well-defined value on $\cV$:  even though $\s$ is branched over $\cV$, so multi-valued on $\cC$, it is single-valued in ``wedge domains" in $\cC$ with edge on $\cV$.  Near points where $\cV$ is a complex submanifold of $\cC$, this is enough to ensure the existence of limiting values of $\s$ \cite{Forstneric:1992}.  As noted below equation \eqref{singularity}, $\cV\setminus\{0\}$ is a complex submanifold of $\cC$ in the rank-2 case we are examining.

We will now argue that $\s$ cannot vanish identically on $\cV$.  In fact, we will show that if $P\in\cV$ is a smooth point of $\cV$,\footnote{I.e., $\cV$ is smooth as a complex subspace of $\cC$ in a neighborhood of $P$} then in any small enough neighborhood $U\subset \cC$ of $P$ the components of $\s$ in $U \cap \cV$ will span a subspace of $V^*$ of dimension at least $2(r-1)$.  This puts strong constraints on the possible EM duality monodromy $M\in\SpDrZ$ around $\cV$ near $P$:  it can be non-trivial only in a single $\Sp(2,\Z)$ subgroup of $\SpDrZ$ involving only the components of $\s$ vanishing in $U \cap \cV$ (see below).

For simplicity and concreteness we will give this argument in the rank-2 case of interest here; the generalization will appear elsewhere \cite{Argyres:2018urp}.  In the vicinity of any point $P\in \cV\setminus \{0\}$, pick good complex coordinates $(u^\perp, u^\parallel)$ vanishing at $P$ such that $\cV$ is given by $u^\perp=0$ in a neighborhood of $P$ and $\del/\del u^\parallel$ is tangent to $\cV$ at $P$.  This is always possible since $\cV\setminus\{0\}$ is a complex submanifold of $\cC$.

Now, distinct points in $\cC$ are necessarily distinct vacua of the UV SCFT, since distinct points have different values of the vevs of local operators in the SCFT.  This means that even though the CB metric is singular (i.e., has non-analytic behavior) at points in $\cV$, the restriction of the CB metric to $u^\perp=0$ must be non-degenerate. For otherwise, if it vanished, there would be no energy cost for fluctuations relating different vacua on the CB, i.e., the distinct vacua on $\cV$ with $u^\perp=0$ but different values of $u^\parallel$ would in fact be the same vacuum: a contradiction.   Thus the $h_{\parallel\bar\parallel}$ component of the CB metric along $\cV$ must be non-zero, giving by \eqref{hmetric}
\begin{align}\label{nondegen}
\vev{\delb_\parallel \sb \, ,\, \del_\parallel\s} \neq 0
\qquad \text{on}\ \cV\setminus\{0\}.
\end{align}
In particular, $\del_\parallel\s \neq 0$ on $\cV$, so $\s$ cannot be identically zero along $\cV$: it must have at least one component which varies with $u^\parallel$.  The same is true of $\sb$, and from \eqref{nondegen} their two components must span a 2-dimensional symplectic subspace of $V^*$.  We will call this symplectic subspace $S_\parallel$, since it is spanned by $\del_\parallel\s$ and $\delb_\parallel \sb$. 

\paragraph{Constraints on the charges which can become massless at $\cV$.}

In the vicinity of a vacuum $P\in\cV\setminus\{0\}$, $\cV$ is described physically as the set of vacua where some charged states become massless.  Denote the set of electric and magnetic $\U(1)^2$ charges of these massless states by $\Phi\subset\Z^4$.  Since charges are integral, $\Phi$ cannot vary as the point $P$ is changed continuously.  Thus $\Phi$ characterizes a whole connected component of $\cV \setminus \{0\}$.\footnote{We already argued in Section \ref{sec2.1} that there are no intervening walls of marginal stability on components of $\cV$ along which $\Phi$ could change discontinuously.}  If $\bq\in\Phi$ then the associated central charge $Z_\bq := \bq^T\s$ vanishes on $\cV$, by definition.  Since the central charge is linear in the charges, if $\bp^T\s$ and $\bq^T\s$ both vanish on $\cV$, then $(\a\bp+\b\bq)^T\s=0$ there as well for arbitrary complex $\a$, $\b$.  Thus algebraically (each component of) $\cV$ is characterized by the complex span of $\Phi$, i.e., a fixed complex linear subspace, $W$, of the complexified charge space $V := \C\otimes\Z^4 \simeq \C^4$.  Note that with respect to the real symplectic structure defined by the charge lattice and its Dirac pairing, complex conjugation maps $W$ to itself.  Thus 
\begin{align}\label{ZonV}
\bw^T\s=\bw^T\sb=0
\quad\text{on $\cV$ for all $\bw\in W$}\, .
\end{align}
This means that at each point of $\cV$, $\s$ takes values only  in the annihilator subspace of $W$.  This is the subspace $W^\text{ann}\subset V^*$ which is the kernel of the dual pairing with $W\subset V$.\footnote{In other words, $W^\text{ann} := \{ \bv\in V^* \, |\, \bw^T\bv=0\ \text{for all}\ \bw\in W\}$.  We do not use the usual notation, ``$W^\perp$", for the annihilator of $W$ since we are reserving $W^\perp$ for the symplectic complement of $W$ in $V$.}

Taking derivatives of \eqref{ZonV} in the $u^\parallel$ direction implies $\bw^T\del_\parallel\s=\bw^T\delb_\parallel\sb=0$ on $\cV$ for all $\bw\in W$.  Thus the 2-dimensional symplectic subspace $S_\parallel\subset V^*$ spanned by $\del_\parallel \s$ and $\delb_\parallel\sb$ on $\cV$ is in the annihilator of $W$:
\begin{align}\label{SinWann}
S_\parallel \subset W^\text{ann}\, .
\end{align}
This implies that $W$ is at most 2-dimensional, and if it is 2-dimensional, then $S_\parallel = W^\text{ann}$ and $W$ is a symplectic subspace of $V$. 

The first two statements are straight forward, and an elementary proof of the last is as follows.  
If $W$ is 2-dimensional, take $\be_j$, $j=1,2$, to be a basis of $W$.  Let $\bs^j$, $j=1,2$, be a basis of $S_\parallel$.  Extend this to a basis of $V^*$, $\bs^a$, $a=1,\ldots,4$, and let the dual basis of $V$ be $\bs_a$.  By definition of the induced Dirac pairing on $V^*$, if $J^{ab}:=\vev{\bs^a,\bs^b}$, then $\vev{\bs_a,\bs_b}= (J^{-1})_{ab}$.  Since $S_\parallel$ is symplectic $J^{12}=\vev{\bs^1,\bs^2}\neq0$  Since $J$ is antisymmetric and non-degenerate, $\vev{\bs_3,\bs_4} = (J^{-1})_{34} = J^{12}/\text{Pf}(J) \neq0$.   Write $\be_i = e_i^a \bs_a$, so $0=\be_i^T \bs^j = e_i^a \bs_a^T \bs^j = e_i^j$ for $i,j=1,2$, since $S_\parallel$ is annihilated by $W$.  Then $\vev{\be_1,\be_2} = e_1^a e_2^b \vev{\bs_a,\bs_b} =  (e_1^3 e_2^4 - e_1^4 e_2^3) \vev{\bs_3,\bs_4}$.  But since $\vev{\bs_3,\bs_4}\neq0$, the  vanishing of right side would imply $\be_1 \parallel \be_2$, contradicting the assumed 2-dimensionality of $W$.

We have shown that the charges of states becoming massless at (a given component of) $\cV$ span at most a 2-dimensional symplectic subspace $W$ of the charge space.  Physically, this simply means that these light states are all charged under only a single low energy $\U(1)$ gauge factor: an appropriate EM duality transformation will set, say, the last two components of  these charge vectors to zero. In this basis these zero components are associated with (dual to) the scalar fluctuations parallel to $\cV$.  

A more invariant way of saying this in the case that $W$ is 2-dimensional is that the charge space splits into two 2-dimensional symplectic subspaces, $V = W \oplus W^\perp$, where $W^\perp$ is the symplectic complement of $W$.   $W$ is the space of electric and magnetic charges of one $\U(1)$ factor, call it ``$\U(1)_\perp$", for which some charged states become massless at $\cV$, while $W^\perp$ is the space of electric and magnetic charges of another, ``$\U(1)_\parallel$", factor for which no charged states become massless at $\cV$.  This basis, $\U(1)_\parallel \times \U(1)_\perp$, of the $\U(1)^2$ vector multiplets reflects the splitting $V^* = S_\parallel \oplus (S_\parallel)^\perp$ of the dual charge space into symplectic subspaces.\footnote{In the case that $W$ is only 1-dimensional, e.g., states carrying only electric charges with respect to one $\U(1)$ become light, the invariant description is a bit different since now $W \subset W^\perp$.  There is no unique choice of a 2-dimensional symplectic subspace containing $W$ which annihilates $S_\parallel$, and so the subspace of charges whose states all remain massive is ambiguous, and can at best be identified with the equivalence classes $W^\perp/W$.  Although a unique $\U(1)_\parallel \times \U(1)_\perp$ decomposition of the vector multiplets is not determined, the $S_\parallel \oplus (S_\parallel)^\perp$ symplectic decomposition of the dual charge space is still defined.}

In summary, charges becoming massless at $\cV$ are charged under some $\U(1)_\perp$ vector multiplet whose scalar $u^\perp$ generates fluctuations $\del/\del u^\perp$ transverse to $\cV$, and are neutral under the $\U(1)_\parallel$ vector multiplet whose scalar fluctuation $\del/\del u^\parallel$ is parallel to $\cV$.  

The EM duality monodromy, $M_\cV$, suffered by $\s$ upon being continued around a small circle linking $\cV$, is particularly simple in this basis.  Since no light states are charged under the $\U(1)_\parallel$ factor, the central charge in all sectors with non-vanishing charges under $\U(1)_\parallel$ will be non-zero at $\cV$.  Call the subspace of $\U(1)_\parallel$ electric and magnetic charges $W^\perp$.  Then since $\bq^T\s\neq0$ on $\cV$ for any $\bq\in W^\perp$, by shrinking $\g$ to $\cV$ we learn that $\bq^T M_\cV\s =\bq^T \s$.  Taking the $u_\parallel$ derivative of this expression then implies that $\bq^T(M_\cV-I)S_\parallel=0$, $\forall \bq\in W^\perp$.  In other words, the $S_\parallel$ symplectic subspace of $V^*$ is an eigenspace of $M_\cV$ with eigenvalue 1. Choosing a basis where the symplectic form is:
\beq
J=\left(
\begin{array}{cc}
\epsilon&0\\
0&\epsilon
\end{array}
\right)\qquad\text{with}\qquad\epsilon=\left(
\begin{array}{cc}
0&1\\
-1&0
\end{array}
\right)
\eeq
and a $\U(1)_\perp \times \U(1)_\parallel$ basis where $V^*$ decomposes into the sum of symplectic subspaces, $V^*=  (S_\parallel)^\perp \oplus S_\parallel$, $M_\cV$ decomposes into $2\times 2$ blocks 
\begin{align}\label{factorize}
M_\cV = \left(
\begin{array}{cc}
M_\perp&D\\
f(D)&I
\end{array}
\right)
\quad\text{with}\quad M_\perp \in \SL(2,\Z),\quad f(D)=\epsilon D^T\epsilon M_\perp, \quad\det{D}=0 .
\end{align} 
We can further massage this expression by using $\Sp(4,\Z)$ matrices which preserves the form \eqref{factorize} to set to zero either a column or a row of $D$, thus we obtain the remarkable constraint that in rank-2, monodromies around complex co-dimension one singularities can be parametrized by only five integers.

\subsection{CB scaling dimensions when $\cV$ is unknotted}\label{sec4.2}

We now apply this understanding to the situation where the only singularities on the CB are the ``unknotted" ones:
\begin{align}\label{}
\cV = \cV_0 \cup \cV_\infty \cup \{0\}\,,
\end{align}
in the notation of Section \ref{cplxscaleact}.  Recall that $\cV_0$ is just the $u=0$ plane and $\cV_\infty$ is the $v=0$ plane in $\cC=\C^2$.  

Call $[\d_0]$ and $[\d_\infty]$ the homotopy classes of simple loops linking $\cV_0$ and $\cV_\infty$, respectively.   Thus, for instance, a representative $\d_0$ loop can be taken to be a circular path around the origin in the $u$ coordinate plane at fixed value of the $v$ coordinate, and similarly for $\d_\infty$ but with the roles of the $u$ and $v$ coordinates reversed.   Let $M_0$ and $M_\infty$ be the EM duality monodromies around $\d_0$ and $\d_\infty$, respectively.

In the vicinity of $\cV_0$, the parallel and transverse coordinates $(u^\parallel,u^\perp)$ are $(v,u)$ respectively
while their roles are reversed around $\cV_\infty$.  Let $S_v$ be the symplectic subspace of $V^*$ spanned by $\del_v\s$ and $\delb_v\sb$ at $\cV_0$.  Then with respect to the symplectic decomposition $V^* = S_v^\perp \odot S_v$, $M_0$ has the block diagonal form 
\begin{align}\label{d0monod}
M_0 = M_u \odot I 
\quad\text{for}\quad M_u\in\Sp(2,\Z)\, ,
\end{align}
where we call it $M_u$ since it is a monodromy in a $u$-plane transverse to $\cV_0$.  Let $\bs_j$, $j=1,2$ be a basis of $S_v$, and let $\bs_j^\perp$ for $j=1,2$ be a basis of $S_v^\perp$ which is a (generalized) eigenbasis of $M_u$. Write $\s$ in this fixed basis as
\begin{align}\label{sdecomp}
\s(u,v) &= \bof_v(u,v) + \bof^\perp_v(u,v) \ , \\[2mm]
\bof_v(u,v) &:= \textstyle{\sum_{j=1,2}}\, f_j(u,v) \, \bs_j 
\quad \in S_v \ ,
\nonumber\\
\bof^\perp_v(u,v) &:= \textstyle{\sum_{j=1,2}}\, f^\perp_j(u,v) \, \bs^\perp_j 
\quad \in S_v^\perp \ ,
\nonumber
\end{align}
for some functions $f_j$, $f^\perp_j$ holomorphic on $\cM = \cC \setminus \cV$.

Now consider a $\d_0$ loop linking $\cV_0$ but at $v=0$, i.e., \emph{inside} the $\cV_\infty$ component of metric singularities.  The homotopy class of such $\d_0\subset \cV_\infty$ loops can be realized by orbits of points under the action of the $\U(1)_R$ isometry acting on the CB.  Explicitly, the $\U(1)_R$ action is the pure phase part of the $\tCs$ action, i.e., it is given by \eqref{scaling-ii} with $\l=\exp\{i\vf\}$ for real $\vf$.   Acting on a point $(u_*,0)\in\C^2$ with this $\U(1)_R$ action gives the image point $(\exp\{i\vf/\D_u\}\, u_*,0)$, so for $\vf\in[0,\vf_*)$ with
\begin{align}\label{phi*d0}
\vf_* := \frac{2\pi}{\D_u}\, ,
\end{align}
this $\U(1)_R$ orbit describes a simple closed $\d_0$ path inside $\cV_\infty$.  

\paragraph{U(1)$_R$ monodromies.}

Such $\U(1)_R$ monodromies have a special property: $\s$ is an eigenvector of such a monodromy with eigenvalue of unit norm.  Since the central charges $Z_\bq = \bq^T \s$ measure masses, the SK section has mass dimension 1.  Therefore, under the complex scaling action \eqref{scaling-ii} the SK section transforms homogeneously with weight one:
\begin{align}\label{sig-scaling}
\s(\l\circ\bu) = \l \, \s(\bu)\, ,
\end{align}
where $\bu=(u,v)$. In particular, under a $\U(1)_R$ action with $\l = \exp\{i\vf\}$, we find that $\s(e^{i\vf}\circ\bu_*) = e^{i\vf} \s(\bu_*)$.  If there is a finite positive smallest value $\vf_*$ of $\vf$ such that the $\U(1)_R$ orbit of the point $\bu_*$ closes, i.e., such that
\begin{align}\label{phistar}
\exp\{i\vf_*\} \circ \bu_* = \bu_*\, ,
\end{align} 
then this orbit describes a closed path, $\g$, in $\cC$ around which we can compute the EM duality monodromy $\mtM_\g$ of $\s$ as
\begin{align}\label{U1RMdef}
\mtM_\g \s(\bu_*) = \s(\exp\{i\vf_*\}\circ \bu_*)\, .
\end{align}
(Recall that we reserve the fancy $\mtM$ for $\U(1)_R$ monodromies.)  It then follows from \eqref{sig-scaling} that
\begin{align}\label{U1RMeigen}
\mtM_\g \s(\bu_*) = \exp\{i\vf_*\}\, \s(\bu_*) \, ,
\end{align}
and so the SK section is an eigenvector of any $\U(1)_R$ monodromy with an eigenvalue of unit norm.

Recall that in addition to having eigenvalues of unit norm, $\SpDtZ$ matrices can also have eigenvalues which lie on the real axis (see appendix \ref{appB0} for details), and so this is a non-trivial constraint on the kinds of $\mtM_\g$ monodromies  that can be realized.

\paragraph{Possible CB dimensions for unknotted singularities.}

We can now apply this to the $\d_0$ monodromy of $\s$ inside the $\cV_\infty$ singularity (which is the $u$ coordinate plane at $v=0$ in $\C^2$).  As we argued in the previous subsection, $\s$ has a well-defined finite limit on $\cV_\infty$, and so the $\d_0$ monodromy, $M_0$, at $v\neq0$ given in \eqref{d0monod} will be equal to the monodromy at $v=0$ as well, by continuity and since the EM duality group $\SpDtZ$ is discrete.  Since the $M_0$ monodromy is also the $\d_0$ monodromy at $v=0$, it is therefore a $\U(1)_R$ monodromy, so we rename it $M_0 \equiv \mtM_0$.  By \eqref{d0monod} it has the block diagonal form
\begin{align}\label{d0U1RM}
\mtM_0 = M_u \odot I
\quad \text{with} \quad M_u\in\Sp(2,\Z)\, ,
\end{align}
with respect to the symplectic decomposition $V^*= S_v^\perp\odot S_v$.  However, because it is a $\U(1)_R$ monodromy, we learn in addition from \eqref{U1RMeigen} and \eqref{phi*d0} that $\mtM_0$ has an eigenvector with eigenvalue
\begin{align}\label{d0U1Rm}
\m = \exp\{2\pi i/\D_u\}\, .
\end{align}
Clearly the eigenvalue of the $I$ block in \eqref{d0U1RM} is $1$.  The possible eigenvalues of unit norm of the $M_u$ block are $\exp\{2\pi i n/k\}$ for $k\in\{1,2,3,4,6\}$ and any integer $n$.  This is a simple property of $\SL(2,\Z)$ matrices, derived in \eqref{conjclass} in appendix \ref{appRk1}.  Since unitarity and the assumption \eqref{freeCBring} imply $\D_u \ge 1$ --- see the discussion above \eqref{unitarity} --- we learn that the possible values of $\D_u$ are
\begin{align}\label{rk1dimlist}
\D_u \in \left\{ 1, \frac65, \frac43, \frac32, 2, 3, 4, 6 \right\}.
\end{align}
Note that this is precisely the set of allowed CB dimensions for rank-1 theories, recorded in Table \ref{table:eps} of appendix \ref{appRk1}.

The argument of the last paragraph applies equally well to the $\cV_\infty$ singularity and the $\d_\infty$ monodromy just by everywhere interchanging the roles of $u$ and $v$, giving the symplectic decomposition $V^* = S_u \odot S_u^\perp$ in which $\mtM_\infty = I \odot M_v$ for some $M_v\in\Sp(2,\Z)$.  But it is not clear yet how the $S_u$ subspace defined at $\cV_\infty$ is related to the $S_v$ subspace defined at $\cV_0$, and so the result is that the possible values of $\D_v$ also lie in the same set appearing in \eqref{rk1dimlist}.  

An immediate consequence of this is that $\D_u$ and $\D_v$ are commensurate (since they are, in fact, rational separately).  Recall that we showed in Section \ref{cplxscaleact} that $\D_u$ and $\D_v$ were commensurate if there were any knotted components of $\cV$.  We have now shown that they are also commensurate when there are no knotted components.  Thus in all cases the CB dimensions are commensurate.  As we will discuss in the next subsection, this implies that the $\U(1)_R$ orbits through \emph{any} point in the CB is closed, and gives a powerful constraint on the possible structure of $\U(1)_R$ monodromies.

Before we explain that, we outline an argument showing that, in fact, the CB geometry with only unknotted singularities necessarily factorizes, and so describes the CB of two decoupled rank-1 SCFTs or IRFTs.  We do not give the full details of the argument, since it is technical in the IRFT case; we do, however, provide the basic analytic ingredients for making the argument in appendix \ref{appanalytic}.

\paragraph{Factorization of the CB geometry for unknotted singularities.}

If $W_0\subset V$ is the subspace spanned by the electric and magnetic charges of states becoming massless at $\cV_0$ as in \eqref{ZonV}, then $S_v \subset W_0^\text{ann}$ by \eqref{SinWann}.  In the case that $W_0$ is 2-dimensional, then, in fact, $S_v = W_0^\text{ann}$, as remarked below equation \eqref{SinWann}.  But that means, by \eqref{ZonV}, that the $\bof^\perp_v$ components of $\s$ in the eigenbasis decomposition \eqref{sdecomp} vanish on $\cV_0$:
\begin{align}\label{f0on0}
\bof^\perp_v(0,v) = 0\, .
\end{align}

Now $\pi_1( \cM = \cC \setminus \cV )$ is very simple in this unknotted setting:  it is generated by loops, $\d_0$ and $\d_\infty$, linking $\cV_0$ and $\cV_\infty$, respectively, which commute: $\d_0\d_\infty =\d_\infty\d_0$.  This can be visualized as in Figure \ref{knot4} without the red knot.\footnote{Indeed, since this a (very) degenerate case of the general torus link, its knot group is given by \eqref{L(1K1)} with the identifications $\g_0=\d_\infty$, $\g_\infty=\d_0$, and the $f_j=1$.}  The EM duality monodromies $\mtM_0$ and $\mtM_\infty$ around $\d_0$ and $\d_\infty$, respectively, therefore commute
\begin{align}\label{Mcomm}
[\mtM_0 , \mtM_\infty]=0\, .
\end{align} 
But since $\mtM_0$ and $\mtM_\infty$ commute, they have common eigenspaces\footnote{In the case where they have generalized eigenspaces, coming from non-trivial Jordan blocks, the subspace corresponding to a sum of blocks of a given eigenvalue of one matrix will split into a sum of Jordan block subspaces of the commuting matrix, even though their generalized eigenvector bases may not coincide.} and since their symplectic structures also have to match, we must have either
\begin{align}\label{factorcases}
\text{(i):}\quad S_u = S_v\, , 
\qquad \text{or\qquad (ii):} \quad S_u = S_v^\perp\, .
\end{align}
The other four possibilities, i.e., that $S_u$ is the span of one $\bs_i\in S_v$ and one $\bs^\perp_j \in S_v^\perp$, cannot be realized because those spans are lagrangian, not symplectic, subspaces of $V^*$.  

In case (i) we have, by the same reasoning that led to \eqref{f0on0}, that $\bof^\perp_v(u,0)=0$ as well.  In this case the only non-vanishing components of $\s$ at $\cV_0$ and $\cV_\infty$ are $\bof_v \in S_u = S_v$.  But these are the eigenspaces of the $I$ factor of both the $\mtM_0$ and $\mtM_\infty$ monodromies.  Therefore $\mtM_0$ and $\mtM_\infty$ must both have eigenvalue $\m=+1$.  This implies by \eqref{d0U1Rm} and its analog for $\D_v$ that $\D_u=\D_v=1$.  But this is a free field theory describing two massless vector multiplets, and so, in fact has no singularities at all.  In other words, in this case the potentially non-trivial $\SL(2,\Z)$ parts of the $\mtM_{0,\infty}$ monodromies are trivial: $M_u=M_v=I$.

Case (ii) is less trivial.  Now the same reasoning implies that in addition to \eqref{f0on0}, we must have
\begin{align}\label{g0oninfty}
\bof_v(u,0) = 0\, .
\end{align}
In this case the non-vanishing components of $\s$ at $\cV_0$ is $\bof_v\in S_v$ and at $\cV_\infty$ is $\bof^\perp_v \in S_u$.  These are now the eigenspaces of the $M_u$ and $M_v$ $\SL(2,\Z)$ factors of the $\mtM_0$ and $\mtM_\infty$ monodromies, respectively.  Therefore, acting on these eigenspaces, the $\U(1)_R$ monodromies are $\mtM_0=M_u$ and $\mtM_\infty=M_v$, inside the $\cV_\infty$ and $\cV_0$ singularities, respectively.  Furthermore, this restricted $\SL(2,\Z)$ monodromy problem in the two singularity components is equivalent to the rank-1 monodromy problem for $\s$ analyzed in appendix \ref{appRk1}.  Thus, we find that
\begin{align}\label{fgbc}
\bof_v(0,v) &= \text{rank-1 $\s(v)$ for elliptic $\SL(2,\Z)$ monodromy $M_v$,}
\nonumber\\
\bof^\perp_v(u,0) &= \text{rank-1 $\s(u)$ for elliptic $\SL(2,\Z)$ monodromy $M_u$.}
\end{align}
Given the boundary conditions \eqref{f0on0}, \eqref{g0oninfty}, and \eqref{fgbc}, it is trivial to perform the analytic continuation to find that $\bof_v(u,v)=\bof_v(v)$ and $\bof_v^\perp(u,v)=\bof_v^\perp(u)$ for all $(u,v)\in\C^2$.  Together with \eqref{sdecomp} and the fact that $\bof_v$ and $\bof_v^\perp$ are valued in symplectic complements, the K\"ahler potential \eqref{Kpot} for this geometry is $K = i \vev{\bar\bof_v(\ub),\bof_v(u)} + i \vev{\bar\bof_v^\perp(\vb),\bof_v^\perp(v)}$, and so the geometry factorizes into a direct product of rank-1 SCFT CB geometries.

This argument made the assumption that the subspaces $W_{0,\infty}$ spanned by the charges of states becoming massless at $\cV_{0,\infty}$, respectively, were both 2-dimensional.  This is equivalent to assuming that there are simultaneously electrically and magnetically charged states becoming massless at each singularity and so that each is described by a rank-1 interacting SCFT, as found above.  

If, instead, one or both of $W_{0,\infty}$ were 1-dimensional, the argument given above for the $\bof_v$ and $\bof^\perp_v$ boundary conditions \eqref{f0on0} and \eqref{g0oninfty} breaks down.  Physically, only electrically charged states become massless at one or both of the singularities, describing rank-1 IR-free theories (IRFTs) instead of SCFTs.  In this case the $M_{u,v}$ $\SL(2,\Z)$ monodromies are of parabolic type, meaning they have non-trivial Jordan blocks, and the behavior of $\s$ near the singularity is more complicated, as outlined at the end of appendix \ref{appRk1}.

This case can be systematically analyzed by solving directly for the analytic structure of $\s$ in the vicinity of a component of $\cV$ in terms of the generalized eigenvector (Jordan block) decomposition of its monodromy.  We record this analytic form for $\s$ in appendix \ref{appanalytic}.  Though we will make no further use of this analytic form in this paper, it will presumably be useful for future efforts to construct all scale-invariant CB geometries by analytic continuation from their boundary values at the locus $\cV$ of metric singularities.

\subsection{Lagrangian eigenspaces of U(1)$_R$ monodromies}\label{sec4.3}

Consider a point, $P_*$, on the CB which is not on either of the unknotted $\tCs$ orbits.  This is a point with coordinates $\bu_* =(u_*,v_*)\in\C^2$ with $u_*\neq0$ and $v_*\neq 0$.  The $\U(1)_R$ orbit through this point is the set $\{ \bu= e^{i\vf}\circ \bu_* , \ \vf\in\R \}$, where the $\tCs$ action, ``$\circ$", is given by \eqref{scaling-ii}.  As long as $\D_u$ and $\D_v$ are commensurate, this orbit forms a closed path.  To see this, define the positive coprime integers $p$ and $q$ by $q/p=\D_u/\D_v$ as we did before in \eqref{Duvtopq}, and define the real number
\begin{align}\label{ratiodef}
s := \frac{\D_u}{q} = \frac{\D_v}{p}\, .
\end{align}
Then the smallest positive value of $\vf$ such that $e^{i\vf} \circ \bu_* = \bu_*$ is easily checked to be $\vf=2\pi/s$.  Thus
\begin{align}\label{gcycle}
\g_{p,q} := 
\{ u = u_* e^{iqs\vf} , v=v_* e^{ips\vf}\, ,
\vf\in [0,2\pi/s) \}\, ,
\end{align}
describes a simple closed path in the CB.  Note that this path is homotopic to the $\U(1)_R$ orbits through other points in a small enough neighborhood of $\bu_*$.

By our argument on $\U(1)_R$ monodromies in the last subsection, \eqref{U1RMeigen} holds:  $\s(\bu_*)$ is an eigenvector of the $\U(1)_R$ monodromy $\mtM_{p,q} \in \SpDtZ$ around $\g_{p,q}$ with an eigenvalue $\m$ of unit norm:
\begin{align}\label{eigeneqn}
\mtM_{p,q} \, \s(\bu_*) = \m \, \s(\bu_*)
\quad\text{with}\quad \m= \exp\{ 2\pi i/s\}\, .
\end{align}
Since $\g_{p,q}$ is homotopic to nearby $\U(1)_R$ orbits, it follows that \eqref{eigeneqn} holds not just at $\bu_*$ but in a whole open neighborhood of $\bu_*$.  Then taking the $\bu$-derivatives of \eqref{eigeneqn} gives
\begin{align}\label{eigeneqn1}
\mtM_{p,q} \, d\s = \m \, d\s
\quad\text{with}\quad \m= \exp\{ 2\pi i/s\}\, , 
\end{align}
in this neighborhood.  Writing $d\s= \del_u\s du + \del_v \s dv$, we see that this means that the vectors $\del_u\s$ and $\del_v\s$ are in the $\m$ eigenspace of $\mtM_{p,q}$.  Recall from the discussion in Section \ref{sec4.0} that regularity of the K\"ahler metric and the SK integrability condition imply that $\del_u\s$ and $\del_v\s$ span a lagrangian subspace of $V^*$.  Thus we learn:
\begin{align}\label{Meigenspace}
\text{The $\m$ eigenspace of $\mtM_{p,q}$ contains a lagrangian subspace.}
\end{align}

This constraint greatly restricts the allowed conjugacy class of the $\mtM_{p,q}\in\SpDtZ$ monodromy.  Appendix \ref{appB0} lists the $\Sp(4,\R)$ conjugacy classes.  Using this list it is a simple matter to find the ones with a unit norm eigenvalue whose eigenspace contains a lagrangian subspace; these are listed in \eqref{Sp4Rclasses}.  It turns out that these are matrices all of whose eigenvalues have unit norm.  Since the $\SpDtZ$ conjugacy classes are subsets of $\Sp(4,\R)$ conjugacy classes, this is also true of all $\SpDtZ$ elements that satisfy \eqref{Meigenspace}.  So even though only a single unit-norm eigenvalue of $\mtM_{p,q}$ is required by virtue of its being a $\U(1)_R$ monodromy, nevertheless:
\begin{align}\label{Meigenvalues}
\text{All of the eigenvalues of $\mtM_{p,q}$ have unit norm.}
\end{align}

\section{CB operator dimensions from U(1)$_\bR$ monodromies}\label{monodromies}

We now combine the constraints on $\U(1)_R$ monodromies derived in the previous sections with some simple topology of the $\U(1)_R$ orbits to derive a finite set of possible scaling dimensions, $\{\D_u,\D_v\}$, for the CB operators.  

First, note that there are three distinct classes of $\U(1)_R$ orbits in $\C^2\setminus\{0\}$.  We have met them all in the last section, but we reproduce them here:
\begin{align}\label{U1Rorbits}
&&\g_0 &:= \biggl\{ &
u &= 0 , & 
v &=v_* e^{ips\vf}, &
\vf &\in \left[0,\frac{2\pi}{ps}\right) &
\biggr\} \, ,
\nonumber\\
&\qquad\qquad&\g_\infty &:= \biggl\{ &
u &= u_* e^{iqs\vf} , & 
v &= 0, &
\vf &\in \left[0,\frac{2\pi}{qs}\right) &
\biggr\} \, , & \qquad \qquad \quad
\\
&&\g_{p,q} &:= \biggl\{ &
u &= u_* e^{iqs\vf} , &
v &= v_* e^{ips\vf}, &
\vf &\in \left[0,\frac{2\pi}{s}\right) &
\biggr\} \, ,
\nonumber
\end{align}
where $u_*$ and $v_*$ are non-zero complex numbers.  Here we are parameterizing, as before, the commensurate CB dimensions by
\begin{align}\label{Duvpqs}
\D_u := qs,
\quad
\D_v := ps,
\quad
p,q\in\N,
\quad
\gcd(p,q)=1,
\quad
s\in\R^+.
\end{align}
$\g_0$, $\g_\infty$, and $\g_{p,q}$ are homotopic to, respectively, the $K_0$, $K_\infty$ unknots, and the $K(p,q)$ torus knot introduced in Section \ref{sec2.3}.  They depend on a choice of base point $P_*=(u_*,v_*)\in\C^2\setminus\{0\}$.  Define
\begin{align}\label{omegadefn}
\w = u_*^p/v_*^q  \in \P^1\, .
\end{align}
It is easy to see that $\g_{p,q}$ is in the knotted orbit $\cV_\w$, while $\g_0$ lies inside the unknotted complex scaling orbit $\cV_0$, and $\g_\infty$ inside $\cV_\infty$.

Consider a general rank-2 SCFT CB, $\cC=\C^2$.  As explained in \ref{topo}, the subvariety, $\cV$, of metric singularities of $\cC$ is a finite union of distinct $\cV_\w$ complex scaling orbits: $\cV = \cup_j \cV_{\w_j}$.  
All $\g_{p,q}$ with $\w \notin \{\w_j\}\cup\{0,\infty\}$ are homotopic in $\cC\setminus\cV$.  This is easy to see since $\w$ takes values in $\P^1$, so we can continuously deform a $\g_{p,q}$ with one value of $\w$ to another by following a path in $\P^1$ that avoids the finite number of $\w_j$ points as well as the $\w=0$ and $\w=\infty$ points.  

Note, however, that deforming $\w$ continuously to 0 or to $\infty$ is not a homotopy since the unknotted $\g_0$ and $\g_\infty$ orbits have a different topology than the $\g_{p,q}$ knots.  This is reflected in the way the periodicity of the $\vf$ coordinate in \eqref{U1Rorbits} jumps discontinuously at $\w=0$ and $\w=\infty$.  In fact, from these periodicities it is easy to see that as $\w\to 0$ or $\infty$, $\g_{p,q}$ is homotopic to a path that traverses $\g_0$ or $\g_\infty$ an integer number of times:
\begin{align}\label{homo1}
\g_{p,q} \sim (\g_0)^p \sim (\g_\infty)^q\,  .
\end{align}

Thus $\g_0$, $\g_\infty$, and $\g_{p,q}$ represent three distinct homotopy equivalence classes of $\U(1)_R$ orbits in $\cM = \cC\setminus \cV$, the manifold of metrically regular points of the CB.  Denote the $\U(1)_R$ monodromies suffered by $\s$ upon continuation around $\g_0$, $\g_\infty$, and $\g_{p,q}$ by $\mtM_0$, $\mtM_\infty$, and $\mtM_{p,q}$, respectively.  Then the unit-norm eigenvalue property of $\U(1)_R$ monodromies \eqref{U1RMeigen} implies that
\begin{align}
\mtM_0 \, \s(0,v) &= \exp(2\pi i/ps) \, \s(0,v)\, ,
\label{Mm0} \\
\mtM_\infty \, \s(u,0) &= \exp(2\pi i/qs) \, \s(u,0)\, ,
\label{Mmi} \\
\mtM_{p,q} \, \s(u,v) &= \exp(2\pi i/s) \, \s(u,v)\, ,
\label{Mmpq} 
\end{align}
for all $(u,v) \in \cC\setminus\cV$ and with $u \neq 0$ and $v\neq 0$.  Also, the homotopy relations \eqref{homo1} imply
\begin{align}\label{homo}
{\mtM_0}^q = {\mtM_\infty}^p = \mtM_{p,q}\, .
\end{align}

As discussed at length in the previous section, the SK section, $\s$, has a finite, nonzero, and continuous limit as it approaches any point of $\cV\setminus\{0\}$, the locus of metric singularities away from the origin (it is not analytic there --- it has branch points --- but its limit is still well-defined).  Thus, in particular, the above statements \eqref{Mm0}--\eqref{Mmi} about the $\mtM_0$ and $\mtM_\infty$ monodromies hold even if the $u=0$ or $v=0$ planes are in the singular locus.

Because the $\mtM_{p,q}$ monodromy applies to $\U(1)_R$ orbits in all of the regular points of the CB minus the $u=0$ and $v=0$ planes, it satisfies the conditions \eqref{Meigenspace} and \eqref{Meigenvalues} derived in the last section, which stated that its $\exp(2\pi i/s)$ eigenspace must be at least two-dimensional and contain a lagrangian subspace.  In appendix \ref{appE0} we derive the list of possible eigenvalues that $\SpDtZ$ matrices satisfying these conditions can have.  In fact, in that appendix we determine the characteristic polynomials of these matrices.  The characteristic polynomials are invariants of the conjugacy classes of $\SpDtZ$, but typically to each characteristic polynomial there can exist many conjugacy classes.  A list of all $\Sp(4,\Z)$ conjugacy classes  with only unit-norm eigenvalues (what we called ``elliptic-elliptic type" in appendix \ref{appB0}) can be extracted from \cite{Eie:1984, Namikawa:1973}; the subset of such conjugacy classes with no non-trivial Jordan blocks is finite.

In the notation for the characteristic polynomials introduced in appendix \ref{appE0}, there are only five which can correspond to matrices with a lagrangian eigenspace: $[1^4]$, $[2^4]$, $[3^2]$, $[4^2]$ and $[6^2]$.  A characteristic polynomial of the form $[N^{\#}]$ has eigenvalues $\exp\{\pm 2\pi i/N\}$.  Comparing this to \eqref{Mmpq} it follows that 
\begin{align}\label{eqn:list1}
\frac{1}{s} = \pm \frac{1}{N} +C
\quad\text{for}\quad N\in\{1,2,3,4,6\} 
\quad\text{and}\quad C\in\Z.
\end{align}
This implies $s$ is rational and therefore the CB dimensions $\D_u$ and $\D_v$ are rational.  However this does not constrain them to lie in a finite set since there is an infinite set of allowed values for $s$, due to the freedom in choosing $C\in\Z$ in \eqref{eqn:list1}.

Because the $\mtM_0$ and $\mtM_\infty$ monodromies only apply to orbits in the $u=0$ and $v=0$ planes, respectively, and not to an open set in the CB, the conditions \eqref{Meigenspace} and \eqref{Meigenvalues}, which were so restrictive for the $\mtM_{p,q}$ monodromy, do not apply.  But because of the homotopy relations \eqref{homo} and because all the eigenvalues of $\mtM_{p,q}$ have unit norm, it follows that all the eigenvalues of $\mtM_0$ and $\mtM_\infty$, not just the one associated with the eigenspace in which $\sigma$ lies, have unit norm.  This allows the classification of their possible characteristic polynomials as products of cyclotomic polynomials.  Using this, in appendix \ref{appE0} we show that the characteristic polynomials of $\mtM_{0,\infty}$ can be one of nineteen possibilities, listed in \eqref{19F}.  This determines the set of possible eigenvalues that these monodromies can have.  

\begin{table}
\centering
$\begin{array}{|c|c|}
\multicolumn{2}{c}{{\text{\bf\quad Possible CB scaling dimensions of rank-2 SCFTs\quad}}}\\
\hline\hline
&\\
\multirow{-2}{*}{$\quad \text{fractional}\quad\,$}&\multirow{-2}{*}{{\large 
$\ \frac{12}{11} \, ,\, \frac{10}{9} \, ,\, \frac{8}{7} \, ,\,
\frac{6}{5} \, ,\, \frac{5}{4} \, ,\, \frac{4}{3} \, ,\, 
\frac{10}{7} \, ,\, \frac{3}{2} \, ,\, \frac{8}{5} \, ,\, 
\frac{5}{3} \, ,\, \frac{12}{7} \, ,\, \frac{12}{5} \, ,\, 
\frac{5}{2} \, ,\, \frac{8}{3} \, ,\, \frac{10}{3} \ $}}\\
\hline
&\\
\multirow{-2}{*}{$\quad \text{integer}\quad\,$}&\multirow{-2}{*}{
$\ 1 \ ,\ 2 \ ,\ 3 \ ,\ 4 \ ,\  5 \ ,\  6 \ ,\  8 \ ,\  10 \ ,\ 12\ $}\\
\hline
\end{array}$
\caption{List of the allowed values of scaling dimensions of CB operators for rank-2 $\cN=2$ SCFTs, with the assumption that the CB chiral ring is freely generated.}
\label{ScaDimList} 
\end{table}

Writing these eigenvalues in the form $\exp(2\pi i B/A)$ where $A>B$, $A,B\in\N$, and gcd$(A,B)=1$ gives a finite list of possible $(A,B)$ pairs (there are 24 possible pairs).  Calling $(A_0,B_0)$ and $(A_\infty,B_\infty)$ the pairs corresponding to the eigenvalues of the $\mtM_0$ and $\mtM_\infty$ monodromies, respectively, we read off from \eqref{Mm0} and \eqref{Mmi} that
\begin{align}\label{expre}
\frac1{ps} = \frac{B_0}{A_0} + C_0\, ,
\quad{\rm and}\quad
\frac1{qs} = \frac{B_\infty}{A_\infty} + C_\infty\, ,
\quad{\rm with}\quad C_\infty, C_0 \in \N.
\end{align} 
The unitarity bounds together with \eqref{Duvpqs} imply the left sides of these equations are less than or equal to one, which in turn implies that $C_0=C_\infty=0$ in \eqref{expre}.  We are therefore left with a finite set of 24 allowed scaling dimensions for $\D_{u,v}$.  The list of allowed values for $\D_u$ and $\D_v$, separated into fractional and integers values, is reported in Table \ref{ScaDimList}, while in Tables \ref{ScaDim}, \ref{ScaDim1}, and \ref{ScaDim2} we collect the details of the monodromy assignments for the different values of $\D_{u,v}$.

It is important to stress that we have not imposed all the constraints implied by our topological arguments.  For instance, we have only listed here the possible set of values either $\D_u$ or $\D_v$ can take. A simultaneous assignment of $\D_u$ and $\D_v$ from this list determines $s$ which then also has to satisfy \eqref{eqn:list1}.  Not all pairs do satisfy this condition: of the 300 possible distinct assignments of $\D_u$ and $\D_v$ from the list of 24 possible values in Table \ref{ScaDimList}, only 244 satisfy this constraint.  

\begin{table}
\centering
$\begin{array}{|c|c|c|c|c|}
\multicolumn{5}{c}{\multirow{-2}{*}{\text{\bf\qquad Rank-2 U(1)$_\bR$ monodromy classes and scaling dimensions (I)\qquad}}}\\
\hline\hline
\quad \mtM_0 \ \text{or}\ \mtM_\infty \quad \, 
&\qquad \D_v \ \text{or}\ \D_u\qquad \,
&\quad  \mtM_{p,q}\quad\, 
&\qquad\qquad s\qquad\qquad\, 
&\ p \ \text{or}\ q\ \, \\
\hline
[1^4]&1&[1^4]&1&1\\\cline{1-5}
[2^4]&2&[1^4]&1&2\\\hline
&&&&\\
\multirow{-2}{*}{$[1^22^2]$}&\multirow{-2}{*}{$2,1$}&\multirow{-2}{*}{$[1^4]$}&\multirow{-2}{*}{$1,\ ${\large $\frac12$}}&\multirow{-2}{*}{$2$}\\\hline
&&&&\\
&&\multirow{-2}{*}{$[3^2]$}&\multirow{-2}{*}{$3,\ ${\large $\frac32$}}&\multirow{-2}{*}{$1$}\\
&&&&\\
\multirow{-4}{*}{$[3^2]$}&\multirow{-4}{*}{$3,\ ${\large $\frac32$}}&\multirow{-2}{*}{$[1^4]$}&\multirow{-2}{*}{$1,\ ${\large $\frac12$}}&\multirow{-2}{*}{$3$}\\\hline
&&&&\\
&&\multirow{-2}{*}{$[4^2]$}&\multirow{-2}{*}{$4,\ ${\large $\frac43$}}&\multirow{-2}{*}{$1$}\\
&&&&\\
&&\multirow{-2}{*}{$[2^4]$}&\multirow{-2}{*}{$2,\ ${\large $\frac23$}}&\multirow{-2}{*}{$2$}\\
&&&&\\
\multirow{-6}{*}{[$4^2$]}&\multirow{-6}{*}{$4,\ ${\large $\frac43$}}&\multirow{-2}{*}{$[1^4]$}&\multirow{-2}{*}{$1,\ ${\large $\frac13$}}&\multirow{-2}{*}{$4$}\\\cline{1-5}
&&&&\\
&&\multirow{-2}{*}{$[6^2]$}&\multirow{-2}{*}{6,\ {\large $\frac65$}}&\multirow{-2}{*}{$1$}\\
&&&&\\
&&\multirow{-2}{*}{$[3^2]$}&\multirow{-2}{*}{{\large$\frac32$},\ {\large$\frac{3}{10}$}\ $\bigg\vert$\ 3,$\ ${\large$\frac35$}}&\multirow{-2}{*}{$4\ \bigg\vert\ 2$}\\
&&&&\\
&&\multirow{-2}{*}{$[2^4]$}&\multirow{-2}{*}{$2,$\ {\large$\frac25$}}&\multirow{-2}{*}{$3$}\\
&&&&\\
\multirow{-8}{*}{[$6^2$]}&\multirow{-8}{*}{6,\ {\large $\frac65$}}&\multirow{-2}{*}{$[1^4]$}&\multirow{-2}{*}{$1,$\ {\large$\frac15$}}&\multirow{-2}{*}{$6$}\\\hline
\end{array}$
\caption{List of the CB operator dimension, $\D_{u,v}$, and $\mtM_{p,q}$ $\U(1)_R$ monodromies that are compatible with a given $\mtM_{0,\infty}$ monodromy.  The last two columns give the values of the $s$, $p$, $q$ parameters which can be realized by simultaneous solutions for both $\D_u$ and $\D_v$.  How to use this information to deduce the allowed pairs of $(\D_u, \D_v)$ values is explained in the text.}
\label{ScaDim} 
\end{table}

\begin{table}
\centering
$\begin{array}{|c|c|c|c|c|}
\multicolumn{5}{c}{\multirow{-2}{*}{\text{\bf\qquad Rank-2 U(1)$_\bR$ monodromy classes and scaling dimensions (II)\qquad}}}\\
\hline\hline
\quad \mtM_0 \ \text{or}\ \mtM_\infty \quad \, 
&\qquad \D_v \ \text{or}\ \D_u\qquad \,
&\quad  \mtM_{p,q}\quad\, 
&\qquad\qquad s\qquad\qquad\, 
&\ p \ \text{or}\ q\ \, \\
\hline
&&&&\\
\multirow{-2}{*}{$[1^23]$}&\multirow{-2}{*}{$3,$\ {\large$\frac32$}, 1}&\multirow{-2}{*}{$[1^4]$}&\multirow{-2}{*}{$1,$\ {\large$\frac13$},\ {\large$\frac12$}}&\multirow{-2}{*}{$3$}\\\cline{1-5}
&&&&\\
\multirow{-2}{*}{$[1^24]$}&\multirow{-2}{*}{$4,$\ {\large$\frac43$}, 1}&\multirow{-2}{*}{$[1^4]$}&\multirow{-2}{*}{$1,$\ {\large$\frac13$},\ {\large$\frac14$}}&\multirow{-2}{*}{$4$}\\\cline{1-5}
&&&&\\
\multirow{-2}{*}{$[1^26]$}&\multirow{-2}{*}{$6,$\ {\large$\frac65$}, 1}&\multirow{-2}{*}{$[1^4]$}&\multirow{-2}{*}{$1,$\ {\large$\frac15$},\ {\large$\frac16$}}&\multirow{-2}{*}{$6$}\\\cline{1-5}
&&&&\\
\multirow{-2}{*}{$[2^23]$}&\multirow{-2}{*}{$3,2,$\ {\large$\frac32$}}&\multirow{-2}{*}{$[1^4]$}&\multirow{-2}{*}{$$\ {\large$\frac12$},\ {\large$\frac13$},\ {\large$\frac14$}}&\multirow{-2}{*}{$6$}\\\cline{1-5}
&&&&\\
\multirow{-2}{*}{$[2^24]$}&\multirow{-2}{*}{$4,2,$\ {\large$\frac43$}}&\multirow{-2}{*}{$[1^4]$}&\multirow{-2}{*}{$1,$\ {\large$\frac12$},\ {\large$\frac13$}}&\multirow{-2}{*}{$4$}\\\hline
&&&&\\
&&\multirow{-2}{*}{$[2^4]$}&\multirow{-2}{*}{$2,$\ {\large$\frac23$},\ {\large$\frac25$}}&\multirow{-2}{*}{$3$}\\
&&&&\\
\multirow{-4}{*}{[$2^26$]}&\multirow{-4}{*}{$6,2,\ ${\large $\frac65$}}&\multirow{-2}{*}{$[1^4]$}&\multirow{-2}{*}{$1,$\ {\large$\frac12$},\ {\large$\frac15$}}&\multirow{-2}{*}{$6$}\\\hline
&&&&\\
\multirow{-2}{*}{$[3\cdot4]$}&\multirow{-2}{*}{$4,3,$\ {\large$\frac32$},\ {\large$\frac43$}}&\multirow{-2}{*}{$[1^4]$}&\multirow{-2}{*}{$$\ {\large$\frac13$},\ {\large$\frac14$},\ {\large$\frac18$},\ {\large$\frac19$}}&\multirow{-2}{*}{$12$}\\\hline
&&&&\\
&&\multirow{-2}{*}{$[3^2]$}&\multirow{-2}{*}{$$\ {\large$\frac32$},\ {\large$\frac43$},\ {\large$\frac34$},\ {\large$\frac3{10}$}\ $\bigg\vert$\ 3,\ {\large$\frac32$},\ {\large$\frac34$},\ {\large$\frac35$}}&\multirow{-2}{*}{$4\ $\bigg\vert$\ 2$}\\
&&&&\\
\multirow{-4}{*}{[$3\cdot6$]}&\multirow{-4}{*}{$6,3,\ ${\large $\frac32,\frac65$}}&\multirow{-2}{*}{$[1^4]$}&\multirow{-2}{*}{$1,$\ {\large$\frac12$},\ {\large$\frac14$},\ {\large$\frac15$}}&\multirow{-2}{*}{$6$}\\\hline
&&&&\\
\multirow{-2}{*}{$[4\cdot6]$}&\multirow{-2}{*}{$6,4,$\ {\large$\frac43$},\ {\large$\frac65$}}&\multirow{-2}{*}{$[1^4]$}&\multirow{-2}{*}{$$\ {\large$\frac12$},\ {\large$\frac13$},\ {\large$\frac19$},\ {\large$\frac1{10}$}}&\multirow{-2}{*}{$12$}\\\hline
\end{array}$
\caption{Continuation of Table \ref{ScaDim}.}
\label{ScaDim1} 
\end{table}

\begin{table}
\centering
$\begin{array}{|c|c|c|c|c|}
\multicolumn{5}{c}{\multirow{-2}{*}{\text{\bf\qquad Rank-2 U(1)$_\bR$ monodromy classes and scaling dimensions (III)\qquad}}}\\
\hline\hline
\quad \mtM_0 \ \text{or}\ \mtM_\infty \quad \, 
&\qquad \D_v \ \text{or}\ \D_u\qquad \,
&\quad  \mtM_{p,q}\quad\, 
&\qquad\qquad s\qquad\qquad\, 
&\ p \ \text{or}\ q\ \, \\
\hline
&&&&\\
\multirow{-2}{*}{$[5]$}&\multirow{-2}{*}{$5,$\ {\large$\frac52$},\ {\large$\frac53$},\ {\large$\frac54$}}&\multirow{-2}{*}{$[1^4]$}&\multirow{-2}{*}{$1,$\ {\large$\frac12$},\ {\large$\frac13$},\ {\large$\frac14$}}&\multirow{-2}{*}{$5$}\\\hline
&&&&\\
&&\multirow{-2}{*}{$[4^2]$}&\multirow{-2}{*}{\ {\large$\frac43$},\ {\large$\frac49$},\ {\large$\frac4{15}$},\ {\large$\frac4{21}$}\ $\bigg\vert$\ 4,\ {\large$\frac23$},\ {\large$\frac45$},\ {\large$\frac47$}}&\multirow{-2}{*}{$6\ $\bigg\vert$\ 2$}\\
&&&&\\
&&\multirow{-2}{*}{$[2^4]$}&\multirow{-2}{*}{$2,$\ {\large$\frac23$},\ {\large$\frac25$},\ {\large$\frac27$}}&\multirow{-2}{*}{$4$}\\
&&&&\\
\multirow{-6}{*}{[$8$]}&\multirow{-6}{*}{$8,$\ {\large$\frac83$},\ {\large$\frac85$},\ {\large$\frac87$}}&\multirow{-2}{*}{$[1^4]$}&\multirow{-2}{*}{$1,$\ {\large$\frac13$},\ {\large$\frac15$},\ {\large$\frac17$}}&\multirow{-2}{*}{$8$}\\\hline
&&&&\\
&&\multirow{-2}{*}{$[2^4]$}&\multirow{-2}{*}{$2,$\ {\large$\frac23$},\ {\large$\frac27$},\ {\large$\frac29$}}&\multirow{-2}{*}{$5$}\\
&&&&\\
\multirow{-4}{*}{[$10$]}&\multirow{-4}{*}{$10$,\ {\large$\frac{10}3$},\ {\large$\frac{10}7$},\ {\large$\frac{10}9$}}&\multirow{-2}{*}{$[1^4]$}&\multirow{-2}{*}{$1,$\ {\large$\frac13$},\ {\large$\frac17$},\ {\large$\frac19$}}&\multirow{-2}{*}{$10$}\\\hline
&&&&\\
&&\multirow{-2}{*}{$[6^2]$}&\multirow{-2}{*}{\ {\large$\frac65$},\ {\large$\frac6{25}$},\ {\large$\frac6{35}$},\ {\large$\frac6{55}$}\ $\bigg\vert$\ 6,\ {\large$\frac65$},\ {\large$\frac67$},\ {\large$\frac6{11}$}}&\multirow{-2}{*}{$10\ $\bigg\vert$\ 2$}\\
&&&&\\
&&\multirow{-2}{*}{$[4^2]$}&\multirow{-2}{*}{\ {\large$\frac43$},\ {\large$\frac4{15}$},\ {\large$\frac4{21}$},\ {\large$\frac4{33}$}\ $\bigg\vert$\ 4,\ {\large$\frac45$},\ {\large$\frac47$},\ {\large$\frac4{11}$}}&\multirow{-2}{*}{$9\ $\bigg\vert$\ 3$}\\
&&&&\\
&&\multirow{-2}{*}{$[3^2]$}&\multirow{-2}{*}{\ {\large$\frac32$},\ {\large$\frac3{10}$},\ {\large$\frac3{14}$},\ {\large$\frac3{22}$}\ $\bigg\vert$\ 3,\ {\large$\frac35$},\ {\large$\frac37$},\ {\large$\frac3{11}$}}&\multirow{-2}{*}{$8\ $\bigg\vert$\ 4$}\\
&&&&\\
&&\multirow{-2}{*}{$[2^4]$}&\multirow{-2}{*}{$2,$\ {\large$\frac25$},\ {\large$\frac27$},\ {\large$\frac2{11}$}}&\multirow{-2}{*}{$6$}\\
&&&&\\
\multirow{-10}{*}{[$12$]}&\multirow{-10}{*}{$12,$\ {\large$\frac{12}5$},\ {\large$\frac{12}7$},\ {\large$\frac{12}{11}$}}&\multirow{-2}{*}{$[1^4]$}&\multirow{-2}{*}{$1$,\ {\large$\frac15$},\ {\large$\frac17$},\ {\large$\frac1{11}$}}&\multirow{-2}{*}{$12$}\\\hline
\end{array}$
\caption{Continuation of Table \ref{ScaDim1}.}
\label{ScaDim2} 
\end{table}

We record in Tables \ref{ScaDim}--\ref{ScaDim2} the detailed monodromy data which characterizes each allowed pair $(\D_u,\D_v)$ of CB operator dimensions.  By scanning the tables one determines the possible eigenvalue classes of the various $\U(1)_R$ monodromies compatible with a given pair of CB dimensions.

As an illustration of how to use the tables, suppose a CB geometry has $\mtM_{p,q}$ monodromy in eigenvalue class $[1^4]$.  Now take a specific instance of the $\D_v$ unknot monodromy, say $\mtM_0 = [2^2 4]$ appearing in the fifth row of Table \ref{ScaDim1}, which has this value of $\mtM_{p,q}$.  Then the possible values of $\D_v$ are $4$, $2$, or $4/3$, with respective $s$ values $1$, $1/2$, or $1/3$, and $p=4$.  In the case where, say, $\D_v=4/3$, thus $s=1/3$ and $p=4$.  Then the possible values of $\D_u$ have to have the same values of $s$, $\mtM_{p,q}$, and a coprime $q$.  These can be determined by scanning the tables.  For instance, $\D_u=4/3$ with $\mtM_\infty=[4^2]$ appearing the bottom line of the fourth row of Table \ref{ScaDim} is not allowed because, though it has $s=1/3$, it has $q=4$ which is not coprime to $p=4$.  On the other hand, $\D_u=5/3$ with $\mtM_\infty = [5]$ appearing in the first row of Table \ref{ScaDim2} is allowed since $q=5$.

Here we will not make any attempt to study the implications of these extra constraints, and leave this analysis for the future.   

Finally, all known examples of rank-2 SCFTs in \cite{Chacaltana:2014jba, Chacaltana:2015bna, Xie:2015rpa, Chacaltana:2016shw, Wang:2016yha, Chacaltana:2017boe, Buican:2017fiq, Distler:2017xba} have CB dimensions which are in the list derived here, though there are entries in our list which do not appear (yet) in any known example.  An earlier attempt at a classification of rank-2 SCFT CBs by one of the authors and collaborators \cite{Argyres:2005pp, Argyres:2005wx} reports some examples with dimensions not appearing in Table \ref{ScaDimList}; however it turns out these conflicting examples are not consistent CB geometries (the geometries in \cite{Argyres:2005pp, Argyres:2005wx} which are incorrect are those with fractional powers of the CB vevs appearing their SW curves; as a result their EM duality monodromies are not in $\SpDtZ$).

\section{Summary and further directions}\label{conclu}

In this paper we took a first step towards generalizing the successful story of the classification of $\cN=2$ SCFTs rank-1 theories \cite{paper1, paper2, allm1602, am1604, Argyres:2016ccharges} to arbitrary ranks.  We illuminated how the special K\"ahler structure, and in particular the $\SpDtZ$ monodromy action, is intricately tied with the globally defined complex scaling action on the CB.  This strongly constrains the scaling dimensions $\D_u$ and $\D_v$ of the CB operators. We obtained the striking result that only a finite list of rational scaling dimensions is allowed for $\D_u$ and $\D_v$.  The allowed values are listed in Table \ref{ScaDimList}.  In particular the maximum allowed mass dimension of rank-2 CB parameters is $\D=12$. 

Using an extension of these arguments, a similar result can be obtained for arbitrary ranks, and will be reported on elsewhere \cite{Argyres:2018urp}.  

Aside from this concrete result on the spectrum of CB scaling dimensions, we have developed a set of tools which we believe will be key to constructing all possible scale invariant rank-2 CB geometries.  Our key results are:  the algebraic description of the possible varieties, $\cV$, of CB singularities in \eqref{singularity}; the computation of the possible topologies of the $\cV\subset\cC$ given in \eqref{L(1K1)}; the factorized description of the local EM duality monodromy $M_\cV$ linking components of $\cV$ in terms of $\Sp(2,\Z)$ matrices given in \eqref{factorize}; the fact that the SK section is an eigenvector of $\U(1)_R$ monodromies with unit-norm eigenvalue \eqref{U1RMeigen}; the lagrangian eigenspace property \eqref{Meigenspace} and fact that all eigenvalues have unit norm \eqref{Meigenvalues} of the generic (knotted) $\U(1)_R$ monodromy; and the interrelations of the three different $\U(1)_R$ monodromies recorded in Tables \ref{ScaDim}--\ref{ScaDim2}.

The next steps towards the goal of constructing all scale-invariant rank-2 CB geometries are likely:
\begin{enumerate}
\item Analyze the implications of the $\U(1)_R$ monodromy conditions found in this paper.  Here we only analyzed the compatibility of the eigenvalues for these matrices, but these conditions imply also that that the associated eigenspaces need to coincide.  Presumably this is a non-trivial constraint which imposes further restriction on the allowed pairs of scaling dimensions $(\D_u,\D_v)$.
\item Investigate the constraints coming from the relationship between the factorized form \eqref{factorize} of monodromies linking single components of $\cV$ and the $\U(1)_R$ monodromies (which, in some sense, link all the components at once).  These monodromies are related by the knot group \eqref{L(1K1)} which reflects the presence of unknots and/or multiple component of the torus links.  In the analysis of this paper, the allowed CB operator dimensions we found only depended on the integers $(p,q)$ characterizing the $\U(1)_R$ orbit but did not depend on the number or type of components in $\cV$.  But the expression for the knot group reflects the existence of all the components of $\cV$ and should be reflected in further constraints on the allowed monodromies and thus on the allowed scaling dimensions.
\end{enumerate}

Two longer-term generalizations of the current project are to extend our considerations to higher rank CBs, and to characterize the mass (or other relevant) deformations of the scale-invariant geometries considered here.  For the higher-rank generalization, one potential technical hurdle is that, to the best of our knowledge, the full classification of non-hyperbolic conjugacy classes of $\Sp(2r,\Z)$ for $r\geq3$ is not known. It is also currently unclear to us whether the full list of these conjugacy classes is actually needed --- the partial results of this paper only required coarser and more easily obtained information about the EM duality group.  While the approach to non-scale-invariant geometries by deformation of scale-invariant ones was fruitful in the rank-1 case \cite{paper1, paper2, allm1602, am1604, Argyres:2016ccharges}, it is already apparent from the structures found in this paper that most tools that worked in rank-1 are not generalizable in a straightforward way to higher ranks.  On the other hand, we are also not aware of any insurmountable obstacle for the implementation of such a program in rank-2.

\acknowledgments 

It is a pleasure to thank Y. L\"u for collaborating in the early stages of the project and for sharing with us useful insights. Furthermore we would like to thank J. Distler, B. Ergun, I. Garc\'ia-Etxebarria, J. Halverson, B. Heidenreich, D. Kulkarni, M. Lotito, D. Regalado and F. Yan for helpful comments and discussions.  PA is supported in part by DOE grant DE-SC0011784 and by Simons Foundation Fellowship 506770.  CL is supported by NSF grant PHY-1620526.  MM is supported by NSF grant PHY-1151392.

\begin{appendix}

\section{Review of rank-1 scale-invariant SK geometries}
\label{appRk1}

\paragraph{Topology.}

By the assumption that the CB chiral ring is freely generated, in the rank-1 case it has a single generator, and therefore $\cC\simeq\C$ as a complex space.  Choose a complex coordinate $u$ on $\cC$ such that a singularity is located at $u=0$.  The complex scale symmetry gives a holomorphic $\tCs$ action on $\cC$ with $u=0$ as a fixed point.  It is a conformal isometry of the metric on $\cM$ coming from the combination of the actions of the $\U(1)_R$ and dilatation generators on the CB.  Thus this action is simply
\begin{align}\label{rank1scaling}
\l\,\circ: u \mapsto \l^{1/\D_u} u, \quad \l \in \tCs,
\end{align}
where $\D_u$ is the mass scaling dimension of $u$.    Unitarity bounds for 4d CFTs plus the assumption that the CB chiral ring is freely generated imply $\D_u \geq 1$.

Since $\cM \simeq \C\backslash \{0\}$, its fundamental group is generated by a path that circles once around $u=0$ counterclockwise, and there is only a single non-trivial monodromy, $M \in \Sp_\D(2,\Z)$, corresponding to analytic continuation along this path.  We can thus describe the special coordinates $\s(u)\in\C^2$ by a holomorphic field on the $u$-plane minus a cut emanating from the origin, so that $\s$ is continuous on $\cM$ except for a ``jump'' by the linear action of $M$ across the cut.

\paragraph{Geometry.}  

Since the central charge has mass dimension 1, it transforms under \eqref{rank1scaling} as $\s(\l^{1/\D_u} u) = \l \s(u)$.  Thus
\begin{align}\label{rk1section}
\s(u)= u^{1/\D_u}  \bpmat \t \\ 1 \epmat,
\quad\text{for some}\quad \t\in\C\, .
\end{align}
Here we have chosen an overall complex constant factor to set the normalization of the bottom component to 1.  The SK integrability condition is trivially satisfied.  The metric on $\cM$ is $ds^2 = (i/2) \vev{\delb_\ub \bar\s,\del_u \s} dud\ub$ where $\vev{\cdot\, ,\cdot}$ is the Dirac pairing.  With \eqref{rk1section} this gives $ds^2 \sim (\Im\,\t) |u^{(1/\D_u) - 1} |^2 du d\ub$.  Positivity and well-definedness of the metric imply $0<\Im\,\t <\infty$.

\paragraph{Duality.}

For rank-1, any Dirac pairing can be written up to a $\GL(2,\Z)$ change of basis, as $\vev{\bp,\bq} = p_i (J_\D)^{ij} q_j$ with $J_\D = \d J$ for some positive integer $\d$, where $J$ is the usual symplectic form $J=(\bsmat 0 & 1 \\ -1 & 0 \esmat)$.  The EM duality group $\Sp_\D(2,\Z)$ are those $M\in \GL(2,\Z)$ such that $M^T J_\D M = J_\D$.  Therefore $\Sp_\D(2,\Z)$ is actually independent of the choice of $\d$:  for all $\d$, $\Sp_\D(2,\Z) = \Sp(2,\Z) = \{ M \in \GL(2,\Z) | M^T J M = J \}$.  Write $M = \left(\bsmat a & b \\ c & d\esmat\right)$, then $M^TJM = J$ becomes simply $ad-bc = 1$. Thus $\Sp(2,\Z) = \SL(2,\Z)$. 

As we follow a path $u_0 \to e^{2\pi i } u_0$, $\s \to M \s$, and therefore $e^{2\pi i/ \D_u}\left(\bsmat 1 \\ \t \esmat\right) = M\left(\bsmat 1 \\ \t \esmat\right)$ for some $M \in \SL(2,\Z)$.  Therefore, $M$ must have an eigenvalue, $\m=\exp(2\pi i/\D_u)$, with $|\m| = 1$.  The characteristic equation of $M$ is $\m^2 - (\Tr\, M)\m +1 = 0$, since $\det M=1$, so both roots have unit norm if $|\Tr\, M| \le 2$; otherwise neither does.  It is easy algebra to list all the conjugacy classes of $\SL(2,\Z)$ satisfying this trace condition:
\begin{align}\label{conjclass}
\Tr\, M &=2 &
&\Rightarrow &
M &\sim T^n &
\m &= +1
\nonumber\\
\Tr\, M &=1 &
&\Rightarrow &
M &\sim ST \ \ \text{or}\ \ (ST)^{-1} &
\m &= e^{\pm i\pi/3}
\nonumber\\
\Tr\, M &=0 &
&\Rightarrow &
M &\sim S \ \ \text{or}\ \ S^{-1} &
\m &= e^{\pm i\pi/2}
\nonumber\\
\Tr\, M &=-1 &
&\Rightarrow &
M &\sim -ST \ \ \text{or}\ \ (-ST)^{-1} &
\m &= e^{\pm 2i\pi/3}
\nonumber\\
\Tr\, M &= -2 &
&\Rightarrow &
M &\sim -T^n &
\m &= -1
\end{align}
Here $T := \left(\bsmat 1 & 1 \\ 0 & 1 \esmat\right)$ and $S := \left(\bsmat 0 & -1 \\ 1 & 0 \esmat\right)$.  In the first and last line $n$ is an integer.  When $n\neq0$ these are called parabolic conjugacy classes.  All the other cases are called elliptic conjugacy classes. 

Since $\m = e^{2\pi i /\D_u}$ and since $\D_u\ge1$ we immediately read off the list of allowed values of $\D_u$, $M$, and $\t$, shown in Table \ref{table:eps}.   The value of $\t$ is determined by solving for the eigenvector of $M$ normalized as in \eqref{rk1section}.  The first seven entries are scale-invariant singular geometries (flat cones, in this case), and correspond to elliptic conjugacy classes. The eighth entry corresponds to an identity monodromy matrix, and therefore to no singularity (a regular point).  The two entries below the dotted line correspond to parabolic conjugacy classes of $\SL(2,\Z)$. 

\begin{table}
\centering
$\begin{array}{|c|c|c|}
\hline
\ \D_u\ \,& M               & \t                   \\
\hline
6           & ST              & e^{i\pi/3}      \\
4           & S                & i                    \\
3           & (-ST)^{-1} & e^{2i\pi/3}    \\
2           & -I               & \ \text{any}\ \t\ \, \\
3/2        & -ST            & e^{2i\pi/3}     \\
4/3        & S^{-1}      & i                     \\
6/5        & (ST)^{-1}  & e^{i\pi/3}      \\
1           & I                & \text{any}\ \t \\
\hdashline
1           & T^n            & i\infty             \\
2           & -T^n           & i\infty             \\
\hline
\end{array}$
\caption{Possible values of $\D_u$, $M$, and $\t$ for rank-1 CB singularities.\label{table:eps}}
\end{table}

For the parabolic classes there is no scale-invariant solution for $\s$ since $\t = i\infty$.  So we should look for solutions by including the leading corrections to scaling. So, e.g., expand
$\s(u) = u + \s_0 u\left(u/\L\right)^{\b_0} + \s_1 u \ln^{\b_1} \left(u/\L\right)$, where the $\b_j$ are 2-component vectors of exponents correlated with the entries of the $\s_j\in\C^2$, and $\L$ is an arbitrary mass scale.  In the $\D_u =1$ parabolic case we look for a solution to $\s(e^{2\pi i}u) = T^n \s(u)$.  We find $\s_0 =(0\ 0)$, $\b_1 = (1\ 0)$ and $\s_1 = ( \frac{n}{2\pi i }\ 0)$.  Thus for the $T^n$ monodromies we find
\begin{align}
\s = u \bpmat 1+\frac{n}{2\pi i}
\ln\left(\frac{u}{\L}\right)\\ 1 \epmat.
\end{align}
For this solution the metric is $ds^2 = -\frac{n}{4\pi}\left\{\text{ln}\left(\frac{u\ub}{\Lambda^2}\right) +2\right\}dud\ub$.  Note that as $|u|\to 0$, ln$(u\ub) \to -\infty$, so the metric is positive-definite in the vicinity of $u=0$ only for $n>0$.  This metric has a mild non-analyticity at $u=0$ with $2\pi$ opening angle there and positive curvature away from $u=0$.  Thus the $T^n$ monodromies for $n \in \Z^+$ give sensible geometries.  They correspond physically to IR-free $\cN=2$ QED, for example with $n$ charge-1 massless hypermultiplets.  $\L$ is the Landau pole.  A similar story goes for the $-T^n$ monodromies.  They give positive definite metrics for $n\in \Z^+$, corresponding to IRFTs such as $\SU(2)$ with $n+4$ massless fundamental hypermultiplets.

\section{Analytic form of the SK section near $\cV \setminus \{0\}$}\label{appanalytic}

Here we record the analytic form of the SK section, $\s$, in the vicinity of any regular point $P$ of $\cV$, the variety of metric singularities in a rank-$r$ CB.  Since $P$ is a regular point of $\cV$, we can pick local complex coordinates $(u^\perp, \bu^\parallel)$ on the CB vanishing at $P$ such that $u^\perp=0$ describes $\cV$ locally and $\bu^\parallel$ are $r-1$ coordinates such that $\del_{\bu^\parallel}$ are tangent to $\cV$ at $P$.  Then $\s$ depends analytically on $\bu^\parallel$, at least in a neighborhood of $P$.  

For instance, in the rank-2 SCFT case, we can describe $u^\perp$ and $u^\parallel$ explicitly and also find the explicit analytic dependence of $\s$ on $u^\parallel$.  This is because in this case the $u^\parallel$-dependence is determined by the complex scale symmetry; it will not be true for ranks greater than 2.

The SK section transforms as in \eqref{sig-scaling} under the
complex scaling action.  Linearizing around $\l=1$, \eqref{sig-scaling} becomes a differential equation with general solution 
\begin{align}\label{abeqn}
\s = v^{1/\D_v}\, \sh(y)\, , \qquad y:=u v^{-\D_u/\D_v}\, ,
\end{align}
with $\sh$ complex analytic in $y$ except at the CB singularities, since $\s$ complex analytic on the CB minus its singularities.   Consider the vicinity of a regular point of $\cV$,  that is of a point $P \in \cV \setminus \{0\}$.  Say $P$ is a point on a $\cV_\w$ component of $\cV$ with $\w\in\P^1$.  Let $\bu_*$ be the coordinates of $P$, so $y_*=u_* v_*^{-\D_u/\D_v} = \w^{1/p}$ (for some choice of the $p$th root).  Expanding $\s$ around $P$ we have from \eqref{abeqn} (with a slight abuse of notation)
\begin{align}\label{sc*}
\s = (v_*+u^\parallel)^{1/\D_v}\, \sh(u^\perp)\, ,
\qquad u^\parallel := v-v_*\, ,
\qquad u^\perp := y-y_*\, .
\end{align}

Returning now to the general-rank case, we will suppress the uninteresting analytic dependence of $\s$ on the $\bu^\parallel$ coordinates, and focus on the interesting non-analyticities in its dependence on $u^\perp$.  For ease of typing, we will from now on write $u$ for $u^\perp$.

By assumption, there is a CB singularity at $u=0$ around which $\s$ suffers an EM duality monodromy, $M\in \SpDrZ$.  Thus upon continuing $\s$ along a closed path $u (\th) = u_0 e^{i\th}$ for $0\le \th\le2\pi$, encircling $u=0$, 
\begin{align}\label{monod*}
\s(u) \to \s(e^{2\pi i} u) = M \,\s(u)\, .
\end{align}
By writing $\s = \sum_j f_j(u) \bv_j$ where $\bv_j$ is a (generalized) eigenbasis of the monodromy matrix $M$, it is simple to determine from \eqref{monod*} the analytic behavior of $f_j(u)$ around $u=0$.  Explicitly, a complex change of basis brings $M$ to Jordan normal form,   
\begin{align}\label{jnf*}
M &\sim \bigoplus_j M_j &
\C^4 &= \bigoplus_j \C^{n_j} &
& \text{where} &
M_j &= \bpmat 
\m_j & \ 1    &            &               &   \\
      & \m_j\  & \ddots  &               &   \\
      &           & \ddots  & \ \ \ddots &   \\
      &           &            & \ \ \m_j    & 1 \\
      &           &            &                & \m_j 
\epmat
\in \GL(n_j,\C)\, ,
\end{align}
where the index $j$ labels the different Jordan blocks each with eigenvalue $\m_j$.  This basis $\{\bv_1^{(j)},\ldots,\bv_{n_j}^{(j)}\}$, unique up to an overall normalization, of each $\C^{n_j}$ subspace thus satisfies
\begin{align}\label{jnf*1}
M \bv_k^{(j)} = \m_j \bv_k^{(j)} + \bv_{k-1}^{(j)}
\qquad \bigl( \ \bv_0^{(j)} := 0\ \bigr)\, .
\end{align}
Writing \eqref{monod*} in this basis then determines the analytic form of $\s(u)$ to be
\begin{align}\label{sch*}
\s(u) & =
\sum_j u^{\,\n_j} g_j(u) \, 
\sum_{k=1}^{n_j} \bv_k^{(j)} (\m_j)^{k-1} \,
\cP_{n_j-k}\left( \frac{\ln u}{2\pi i} \right)\, , 
\end{align}
where $g_j(u)\in\C$ is analytic in $u$ in a neighborhood of $u=0$, $\n_j$ is defined in terms of $\m_j$ by
\begin{align}\label{nudefn}
\exp\{2\pi i \n_j\} = \m_j
\quad\text{with}\quad 0 \le \Re\,\n_j < 1\, ,
\end{align}
and the $\cP_\ell$ are degree $\ell$ polynomials obeying the recursion relation
\begin{align}\label{Precurs}
\cP_\ell(x+1) = \cP_\ell(x) + \cP_{\ell{-}1}(x)\, .
\end{align}

The $g_j$ are taken to be analytic around $u=0$, and in particular to not have any poles, because $\s(u)$ cannot diverge as $u\to 0$ (this was argued in Section \ref{sec4.1}).   If the $j$th Jordan block has both $\Re\,\n_j=0$ and non-constant polynomial dependence on $\ln u$, then finiteness of $\s(u)$ as $u\to0$ implies the stronger condition that $g_j(u)$ must vanish as $u\to0$.

The recursion relation \eqref{Precurs} does not determine the constant term of each polynomial, and these constants can be chosen independently for each Jordan block.  If we define $c_\ell := \ell!\, \cP_\ell(0)$, then the first six polynomials are
\begin{align}\label{}
\cP_0 &= c_0\nonumber\\
\cP_1 &= c_0 x + c_1\nonumber\\
2!\,\cP_2 &= c_0 x^2 + (2 c_1 - c_0) x + c_2 \nonumber\\
3!\,\cP_3 &= c_0 x^3 + (3 c_1 - 3 c_0) x^2 
+ (3 c_2 - 3 c_1 + 2 c_0) x + c_3 .\nonumber\\
4!\,\cP_4 &= c_0 x^4 + (4 c_1 - 6 c_0) x^3 
+ (6 c_2 - 12 c_1 + 11 c_0) x^2 
+ (4 c_3 - 6 c_2 + 8 c_1 - 6 c_0) x
+ c_4 \nonumber\\
5!\,\cP_5 &= c_0 x^5 + (5 c_1 - 10 c_0) x^4 
+ (10 c_2 - 30 c_1 + 35 c_0) x^3 
+ (10 c_3 - 30 c_2 + 55 c_1 - 50 c_0) x^2
\nonumber\\
& \qquad\qquad \text{} 
+ (5 c_4 - 10 c_3 + 20 c_2 - 30 c_1 + 24 c_0) x
+ c_5\, .\nonumber
\end{align}

The main properties to take away from \eqref{sch*} are that: the eigenvalue $\m_j$ of each Jordan block determines the leading (fractional) powers, $u^{\n_j}$, appearing in $\s$; a Jordan block of size $n_j$ will contribute logarithms in $u$ up to order $\ln^{n_j-1}$; unless a Jordan block has eigenvalue $\m_j=1$, its contribution to $\s$ will vanish at $u=0$; and, if $\m_j=1$ for a Jordan block with $n_j>1$, then its contribution to $\s(0)$ is non-zero only if only its 1-eigenvector contributes (i.e., all the $c_j$ coefficients in the logarithmic polynomials vanish except for $c_{n_j-1}$).

There are further interesting constraints on \eqref{sch*} that come from incorporating the simple factorized form of the $\SpDrZ$ linking monodromy found in \eqref{factorize}, with the properties of $\SL(2,\Z)$ conjugacy classes described in appendix \ref{appRk1}, the conditions \eqref{positivity} for the positivity of the K\"ahler metric near $\cV$, and the conditions \eqref{nondegen} for the nondegeneracy of the metric components parallel to $\cV$ at $P$.

\section{Sp(4,$\,\R$) conjugacy classes}\label{appB0}

Here we summarize following \cite{Freitas:2004} the conjugacy classes of $\Sp(4,\R)$ and some of their properties.  We then use this knowledge to deduce in which of those conjugacy classes an $\mtM_{p,q}$ EM duality monodromy associated to an $\U(1)_R$ orbit (as described in Section \ref{sec4.3}) can appear.

Even though $\Sp(4,\R)$ is not the EM duality group, $\SpDtZ$, we discuss it here because the description of its conjugacy classes is substantially easier than that of $\SpDtZ$.  Since $\SpDtZ$ is a subgroup of $\Sp(4,\R)$ (as explained in appendix \ref{appE0}), the conjugacy classes of $\SpDtZ$ are subsets of the conjugacy classes of $\Sp(4,\R)$, and that turns out to provide enough information for our purposes. 

\paragraph{Generalized eigenvectors and Jordan blocks.}

First, recall the definition of an $\ell$-generalized eigenvector, or $\ell$-eigenvector for short.  An $\ell$-eigenvector, $\bv_\ell$, with eigenvalue $\l$ of a square matrix $M$ is a non-zero vector for which $(M-\l I)^\ell \bv_\ell=0$ for some positive integer $\ell$, but not for $\ell-1$.  If $\ell=1$, then it is a regular eigenvector.  If $\ell>1$, then $\bv_{\ell-m}:=(M-\l I)^m \bv_\ell$ for $m<\ell$ is an $(\ell-m)$-generalized eigenvector.  Thus if there is an $\ell$-generalized eigenvector, the associated eigenvalue must have multiplicity at least equal to $\ell$.  A series of such $\ell$-generalized eigenvectors with eigenvalue $\l$ and $1\le \ell \le n$ correspond to an $n\times n$ Jordan block when $M$ is put in Jordan normal form.  The matrix for an $n\times n$ Jordan block is shown in \eqref{jnf*}.

We define the \emph{generalized eigenspace} with eigenvalue $\l$ to be the direct sum of the spaces of all Jordan blocks with eigenvalue $\l$.

\paragraph{Properties of Sp(2r,$\,\R$).}

The following properties are true for $\Sp(2r,\R)$ for all $r$, and so for the case of interest here, $r=2$.  They are explained in standard texts on symplectic geometry.  $\Sp(2r,\R)$ is the group of $M\in\GL(2r,\R)$ such that $M^T J M=J$ for $J$ a non-degenerate skew-symmetric matrix.  We can choose a basis of $\R^{2r}$ so that 
\begin{align}\label{Jsymp}
J = \bpmat 0&-I\\ I&0\epmat, 
\end{align}
where $I$ is the $r\times r$ identity matrix.  It follows that if $M=(\bsmat A&B\\ C&D\esmat)\in\Sp(2r,\R)$ with $A,B,C,D$ $r\times r$ real matrices, then
\begin{align}\label{GDprop1}
A^T C=C^T A\, , \quad
B^T D = D^T B\, , \quad
A^T D -  C^T B = I\, .
\end{align}
Some basic but non-trivial properties of $M\in\Sp(2r,\R)$ are that $M$ is similar to $M^{-1}$, $\det(M)=1$, the eigenvalues of $M$ occur in reciprocal pairs, and complex eigenvalues have unit norm.  Thus the set of eigenvalues of any $M\in\Sp(4,\R)$ are always of the form $\{\l_1,\l_2,\l_2^{-1},\l_1^{-1}\}$,with each $\l_i$ either complex of norm one or non-zero real.  

Another basic property involves the symplectic-orthogonality of generalized eigenvectors.  Define the symplectic pairing of two vectors in $\C^{2r}$ by $\vev{\bu,\bv} = \bu^T J \bv$.  If $\bu$ and $\bv$ are $1$-eigenvectors of $M\in\Sp(2r,\R)$ with eigenvalues $\l$ and $\m$, respectively, then $\vev{\bu,\bv} =\vev{M\bu,M\bv} = \l\m \vev{\bu,\bv}$.  It follows that $\vev{\bu,\bv}=0$ unless $\l\m=1$, i.e., eigenvectors of non-reciprocal eigenvalues are symplectic-orthogonal.  This property generalizes to the statement that whole generalized eigenspaces of non-reciprocal eigenvalues are symplectic-orthogonal. 

Recall some definitions from symplectic geometry.  If $W$ is a linear subspace of $\C^{2r}$, the \emph{symplectic complement} of $W$ is the subspace $W^\perp := \{\bv\in \C^{2r} | \vev{\bv,\bw}=0 \ \text{for all}\ \bw\in W\}$.  It satisfies $(W^\perp)^\perp =W$ and $\dim W+\dim W^\perp =2r$.  Then $W$ is \emph{symplectic} if $W^\perp\cap W = \{0\}$.  This is true if and only if $\vev{\cdot,\cdot}$ restricts to a nondegenerate form on $W$.  Thus a symplectic subspace is a symplectic vector space in its own right.  $W$ is \emph{isotropic} if $W \subseteq W^\perp$. This is true if and only if $\vev{\cdot,\cdot}$ restricts to $0$ on $W$, i.e., if and only if all vectors in $W$ are symplectic orthogonal. 
Finally, $W$ is \emph{lagrangian} if $W = W^\perp$.  A lagrangian subspace is an isotropic one whose dimension is $r$.  Every isotropic subspace can be extended to a lagrangian one.

Then the symplectic orthogonality of generalized eigenspaces implies that if $W_\l$ is the generalized eigenspace associated with eigenvalue $\l$ then $W_\l$ is isotropic if $\l\neq\pm1$, $W_\l \oplus W_{1/\l}$ is symplectic, and $W_{\pm1}$ are each symplectic.  

\paragraph{Sp(2,$\,\R$) conjugacy classes.}

Next, recall the structure of $\Sp(2,\R) \simeq \SL(2,\R)$ conjugacy classes.  An $M\in\Sp(2,\R)$ has eigenvalues $\{\l,\l^{-1}\}$ with $\l \in \C^*$ and either $|\l|=1$ or $\l\in\R$.  The matrices can be divided into three sets: ``hyperbolic" if $|\l|\neq1$ in which case it is similar to a diagonal matrix, ``parabolic" if $\l=\pm1$ and $M$ is similar to a $2\times2$ Jordan block form (i.e., has a $1$-eigenvector and a $2$-eigenvector), and ``elliptic" if $|\l|=1$ and is similar over the complex numbers to a diagonal matrix (i.e., has two $1$-eigenvectors).  Note that $M=\pm I$ are special cases of the elliptic class.

The eigenvalues together with whether in the case of $\l=\pm1$ there is a generalized eigenvector or not gives a classification of all the $\Sp(2,\R)$ conjugacy classes.  The $\Sp(2,\Z)\simeq \SL(2,\Z)\subset \Sp(2,\R)$ conjugacy classes are a refinement of these conjugacy classes.  In particular, only eigenvalues of the form $e^{i\pi k/3}$ or $e^{i\pi k/2}$ are allowed in the elliptic cases and each value (together with its inverse) corresponds to at most two separate conjugacy classes; the two parabolic classes corresponding to $\l=\pm1$ in the real case each split into an infinite series of conjugacy classes in the integer case; and only rational eigenvalues are allowed in the hyperbolic cases, and there is a complicated pattern of how many conjugacy classes correspond to a given eigenvalue.

\paragraph{Sp(4,$\,\R$) conjugacy classes.}

The $\Sp(4,\R)$ conjugacy classes have a similar, though inevitably more complicated, description in terms of their eigenvalues and whether or not there are $\ell$-generalized eigenvectors than in the $\Sp(2,\R)$ case.   We will adapt the hyperbolic/elliptic nomenclature of the $\Sp(2,\R)$ case to this case by classifying $M$ as: ``hyperbolic-hyperbolic" (HH) if $|\l_i|\neq1$ for both $i=1$ and $2$; ``hyperbolic-elliptic" (HE) if $|\l_1|\neq1$ and $|\l_2|=1$; and ``elliptic-elliptic" (EE) if $|\l_i|=1$ for both $i=1$ and $2$.  We further subdivide these classes by the size of their Jordan blocks when put in Jordan normal form by a complex change of basis.  We will list these sizes as subscripts when they are larger than one:  these are the analogs of the parabolic-type elements of $\Sp(2,\R)$.  One then easily finds that only the following cases are allowed in $\Sp(4,\R)$:
\begin{align}\label{}
& (HH) \qquad (HH)_{2,2} 
\nonumber\\
& (HE) \qquad (HE)_{2} 
\\
& (EE) \qquad (EE)_{2} \qquad (EE)_{2,2} \qquad (EE)_{4}\, .
\nonumber
\end{align}

It takes considerably more work \cite{Freitas:2004} to describe the different $\Sp(4,\R)$ conjugacy classes realizing these cases.  

Given the $2\times 2$ block structure imposed by the choice \eqref{Jsymp} of $J$, there are three useful ways of combining $2\times2$ matrices into $4\times4$ matrices:  the usual direct sum, $\oplus$, what we will call the upper direct sums, $\oslash_\a$ with $\a=\pm1$, and the interlaced direct sum, $\odot$.  If $A=\left(\bsmat a&b\\ c&d\esmat\right)$ and $B=\left(\bsmat p&q\\ r&s\esmat\right)$, then these sums are defined by
\begin{align}\label{}
A\oplus B &:=
\bpmat a & b &  & \\ 
           c & d &  & \\
            &  & p & q\\
            &  & r & s\epmat, &
A\oslash_\a B &:=
\bpmat a & b &  & \\ 
           c & d &  & \\
            & \a & p & q\\
            &  & r & s\epmat, &
A\odot B &:=
\bpmat a &  & b & \\ 
            & p &  & q\\
           c &  & d & \\
            & r &  & s\epmat,
\end{align}
where $\a =\pm1$ and the empty entries are all 0.  The interlaced sum, $\odot$, is the one that respects the symplectic structure:  $A\odot B$ is in $\Sp(4,\R)$ if and only if $A$ and $B$ are in $\Sp(2,\R)$.  This is because $J= J_2\odot J_2$ where $J_2$ is the $2\times2$ symplectic structure.  It follows easily from \eqref{GDprop1} that $A\oplus B$ is in $\Sp(4,\R)$ if and only if $B=A^{-T}$, and $A\oslash_\a B$ is in $\Sp(4,\R)$ if and only if $B=A^{-T}$ and the upper left entry of $A$ vanishes.

Next, define the following six types of $2\times2$ matrices:
\begin{align}\label{}
H_a &= \bpmat a & 0\\ 0 & 1/a\epmat, &
a &\in \R \ \text{and}\ a\neq0,\pm1,
\nonumber\\
\til H_{a\a} &= \bpmat a & 0\\ \a & a\epmat, &
a &\in \R \ \text{and}\ a\neq0,\pm1,
\quad \a \in \{\pm1\},
\nonumber\\
P_{\a\b} &= \bpmat \a&0\\ \b&\a\epmat, &
\a,\b &\in \{\pm1\},
\nonumber\\
\til P_{\a\b} &= \bpmat 0&\b\\ -\b&2\a\epmat, &
\a,\b &\in \{\pm1\},
\\
E_\th &= \bpmat \cos\th & -\sin\th\\ \sin\th & \cos\th \epmat, &
0 &\le \th \le \pi\, ,
\nonumber\\
\til E_\th &= \bpmat 0 & 1\\ -1 & 2\cos\th \epmat\, , &
0 &<\th<\pi \, .
\nonumber
\end{align}
Note that $H_a$, $P_{\a\b}$, and $E_\th$ are representatives of the hyperbolic, parabolic, and elliptic $\Sp(2,\R)$ conjugacy classes, and that $\til P_{\a\b}$ and $\til E_\th$ are conjugate to $P_{\a\b}$ and $E_\th$, respectively, in $\Sp(2,\R)$.  Note also that $E_0 = - E_\pi = I$.

Then representatives of all the $\Sp(4,\R)$ conjugacy classes are \cite{Freitas:2004}
\begin{align}\label{coarselist}
M&\in (HH)&  &\Rightarrow &
M &\sim H_a \odot H_b,
\nonumber\\
M&\in (HH)_{2,2}& &\Rightarrow &
M &\sim \til H_{a\a} \oplus \til H_{a\a}^{-T},
\nonumber\\
M&\in (HE)& &\Rightarrow &
M &\sim H_a \odot E_\th,
\nonumber\\
M&\in (HE)_2& &\Rightarrow &
M &\sim H_a \odot P_{\a\b},
\\
M&\in (EE)& &\Rightarrow &
M &\sim E_\th \odot E_\psi\quad \text{or}\quad \til E_\th\oplus\til E_\th^{-T},
\nonumber\\
M&\in (EE)_2& &\Rightarrow &
M &\sim E_\th \odot P_{\a\b} ,
\nonumber\\
M&\in (EE)_{2,2}& &\Rightarrow &
M &\sim P_{\a\b} \odot P_{\g\d}\quad \text{or}\quad 
\til P_{\a\b}\oplus\til P_{\a\b}^{-T}\quad \text{or}\quad 
\til E_\th \oslash_\a \til E_\th^{-T} ,
\nonumber\\
M&\in (EE)_4& &\Rightarrow &
M &\sim \til P_{\a\b} \oslash_\g \til P_{\a\b}^{-T} .
\nonumber
\end{align}
Note that $A\odot B \sim B \odot A$, and similarly for $\oplus$ and $\oslash$.

\paragraph{Sp(4,$\R$) conjugacy classes with lagrangian 1-eigenspaces.}

In Sections \ref{sec4.2} and \ref{sec4.3} we showed that the EM duality ``knot" monodromies, $\mtM_{p,q}$, must be an element of $\Sp(4,\R)$ with an eigenvalue of unit norm.  From our discussion above, this means the monodromy cannot be of any of the $(HH)$-types shown in the first two lines of \eqref{coarselist}.

In addition, we showed that the unit-norm eigenvalue must have a 1-eigenspace of dimension 2 or greater.   It is not hard to read off from \eqref{coarselist} that the only possible conjugacy classes with this property are
\begin{align}\label{coarselist2}
M&\in (HE)& &\Rightarrow &
M &\sim H_a \odot E_0
\quad \text{or}\quad
H_a \odot E_\pi ,
\nonumber\\
M&\in (EE)& &\Rightarrow &
M &\sim 
E_0 \odot E_\th
\quad \text{or}\quad
E_\pi \odot E_\th
\quad \text{or}\quad
E_\th \odot E_\th 
\quad \text{or}\quad 
\til E_\th\oplus\til E_\th^{-T},
\nonumber\\
M&\in (EE)_2& &\Rightarrow &
M &\sim E_0 \odot P_\a 
\quad \text{or}\quad
E_\pi \odot P_\a,
\nonumber\\
M&\in (EE)_{2,2}& &\Rightarrow &
M &\sim P_\a \odot P_\a
\quad \text{or}\quad 
\til P_\a\oplus\til P_\a^{-T} .
\end{align}

Furthermore, we also showed in Section \ref{sec4.3} that if the 1-eigenspace is 2-dimensional it must be lagrangian, i.e. for any basis $\{\l_1,\l_2\}$ we require $\vev{\l_1,\l_2}=0$.  We check this for the above list.  First, for the conjugacy classes with representatives given by interlaced sums, $\odot$, if both eigenvectors are from the same summand, then they will have $\vev{\l_1,\l_2}\neq 0$.  So only interlaced sums for which eigenvectors for the same eigenvalue come from both sides of the interlaced sum can give lagrangian eigenspaces.  For the remaining direct sum cases, it is easy to check that for $M = \til E_\th\oplus\til E_\th^{-T}$, $\l_1^T=(0\ 0\ -e^{i\th}\ 1)$ and $\l_2^T=(e^{-i\th}\ 1\ 0\ 0)$ are an eigenbasis of the $e^{i\th}$ eigenspace satisfying $\vev{\l_1,\l_2}=0$; and for $M=\til P_\a\oplus\til P_\a^{-T}$, a basis of the $\a$ eigenspace is $\l_1^T=(0\ 0\ -\a\ 1)$ and $\l_2^T=(\a\ 1\ 0\ 0)$ which also satisfies $\vev{\l_1,\l_2}=0$.  Therefore the list of $Sp(4,\R)$ conjugacy classes of ``knot" monodromies that can appear in scale-invariant singularities are:
\begin{align}\label{Sp4Rclasses}
M&\in (EE)& &\Rightarrow &
M &\sim E_\th \odot E_\th
\qquad \text{or}\qquad 
\til E_\th\oplus\til E_\th^{-T},
\nonumber\\
M&\in (EE)_2& &\Rightarrow &
M &\sim E_0 \odot P_1 
\qquad \text{or}\qquad
E_\pi \odot P_{-1},
\\
M&\in (EE)_{2,2}& &\Rightarrow &
M &\sim P_\a \odot P_\a
\qquad \text{or}\qquad 
\til P_\a\oplus\til P_\a^{-T} .
\nonumber
\end{align}
Note that all these conjugacy classes are of ``elliptic-elliptic" type, so, in particular, only have eigenvalues on the unit circle in the complex plane.

\section{Sp$_{\D}$(4,$\,\Z$) characteristic polynomials}\label{appE0}

\paragraph{Properties of Sp$_{\D}$(4,$\,\Z$).}

The EM duality group, $\SpDrZ$, is the subgroup of $\GL(2r,\Z)$ preserving the Dirac pairing on the charge lattice.  If we write $\vev{\bp , \bq} := p_i (J_\D)^{ij} q_j$ for $\bp,\bq\in\Z^{2r}$, then $\SpDrZ$ is defined to be the set of $M \in \GL(2r,\Z)$ such that $M^T J_\D M = J_\D$.  By a $\GL(2r,\Z)$ change of basis any non-degenerate, integral, skew-symmetric quadratic form can be put in the form $J_\D = \left(\bsmat 0 & -\D \\ \D & 0 \esmat\right)$ where $\D$ is a diagonal $r\times r$ matrix $\D=\text{diag}\{\d_1,\d_2,\ldots,\d_r\}$ with the $\d_i$ positive integers such that $\d_i | \d_{i+1}$.  The set $\{\d_i\}$ is a $\GL(2r,\Z)$-invariant characterization of the form.  Sometimes $J_\D$ is called a polarization.   If $\D = I_r$ then $J_\D$ is a principal polarization, and $\SpDrZ = \Sp(2r,\Z)$, the ``usual" EM duality group.  If any of the ratios $\d_{i+1}/\d_i$ are not perfect squares, then $\SpDrZ$ is not isomorphic to $\Sp(2r,\Z)$ as a group.  However, all pairings can be brought to principal form within $\GL(2r,\R)$, so all $\SpDrZ$ are isomorphic to a subgroup of $\Sp(2r,\R)$.  Since the latter fact together with the fact that $\SpDrZ\subset\GL(2r,\Z)$ are the only facts we will use about $\SpDrZ$ in this paper, the distinction between $\SpDrZ$ and the more familiar $\Sp(2r,\Z)$ EM duality group will not play any role.

\paragraph{Possible eigenvalues of elliptic-elliptic elements of Sp$_{\D}$(4,$\,\Z$).}

Following an argument from \cite{Eie:1984} we can easily determine the possible eigenvalues of type $(EE)_n$ elements $M\in \SpDtZ$.  These have eigenvalues of unit norm.  Their characteristic polynomials, $P_M(x)$, have integer coefficients since $M\in\GL(4,\Z)$ is a matrix with integer entries.  

Polynomials irreducible over the integers whose roots have unit norm and have integer coefficients are the cyclotomic polynomials
\begin{align}\label{}
\Phi_n(x) = \prod_{\gcd(m,n)=1} \left(x-e^{2\pi i m/n}\right)\, ,
\end{align}
and satisfy 
degree$(\Phi_n)=\vf(n)=n\prod_{\text{primes}\ p|n}(1-p^{-1})$ which is Euler's totient function, and counts the number of primitive $n$th roots of unity.

Since the degree of $P_M(x)$ is 4, it can can only be a product of $\Phi_n$ of degrees less than 4.  If $n$ has a prime divisor greater than 5 or if it has more than two distinct prime divisors then $\vf(n)>4$.  So the only possible $\Phi_n$ are in the list $n \in \{1,2,3,4,5,6,8,10,12\}$, which have degrees $d_n \in \{1,1,2,2,4,2,4,4,  4\}$, respectively.  From this and the fact that $\Sp(2r,\R)$ eigenvalues always appear in reciprocal pairs, we can read off the 19 possible characteristic polynomials of type $(EE)_n$ elements:
\begin{align}\label{19F}
&[1^4], &
&[2^4], 
\nonumber\\
&[1^2 2^2], &
&[3^2], &
&[4^2], &
&[6^2], 
\\
&[1^2 3], &
&[1^2 4], &
&[1^2 6], & 
&[2^2 3], &
&[2^2 4], &
&[2^2 6], 
\nonumber\\
&[3\cdot4], &
&[3\cdot6], &
&[4\cdot6], &
&[5], &
&[8], & 
&[10], &
&[12],
\nonumber
\end{align}
where we have introduced the notation
\begin{align}\label{}
\bigl[\textstyle{\prod_i} n_i^{r_i}\bigr] := \prod_i (\F_{n_i})^{r_i}\, .
\end{align}
We will also use the symbol $[X]$ to denote the set of elements $M\in\SpDtZ$ whose characteristic polynomials are $[X]$.  Note that if $M\in[X]$, then any $M'$ conjugate to $M$ is also in $[X]$, so $[X]$ is a union of conjugacy classes.
The eigenvalues of any $M\in[X]$ are simply read off as the primitive $n_i$-th roots of unity each with multiplicity $r_i$.  

Note that dimensions of the generalized eigenspaces are $\{4\}$ for the entries in the first line of \eqref{19F}, $\{2,2\}$ for those in the second line, $\{2,1,1\}$ for those in the third, and $\{1,1,1,1\}$ for the last line.

\paragraph{Possible conjugacy classes of Sp$_{\D}$(4,$\,\Z$) with lagrangian 1-eigenspaces.}

By comparing with \eqref{Sp4Rclasses} we determine which $\Sp(4,\R)$ conjugacy classes with a lagrangian 1-eigenspace can occur in $\SpDtZ$ and what are their characteristic polynomials.  
\begin{align}\label{}
M & \in(EE):&
E_0\odot E_0 &\in [1^4] &
E_\pi\odot E_\pi &\in [2^4]
\nonumber\\
&&
E_{2\pi/3}\odot E_{2\pi/3} 
&\in [3^2] &
E_{\pi/2}\odot E_{\pi/2} 
&\in [4^2] &
E_{\pi/3}\odot E_{\pi/3} 
&\in [6^2] 
\nonumber\\
&&
\til E_{2\pi/3}\oplus \til E_{2\pi/3}^{-T}
&\in [3^2] &
\til E_{\pi/2}\oplus \til E_{\pi/2}^{-T}
&\in [4^2] &
\til E_{\pi/3}\oplus \til E_{\pi/3}^{-T}
&\in [6^2] .
\nonumber\\
M & \in(EE)_2 :&
E_0\odot P_1 &\in [1^4] &
E_\pi\odot P_{-1} &\in [2^4]
\nonumber\\
M & \in(EE)_{2,2}: &
P_1\odot P_1 &\in [1^4] &
P_{-1}\odot P_{-1} &\in [2^4]
\nonumber\\
&& 
\til P_1\oplus \til P_1^{-T} &\in [1^4] &
\til P_{-1}\oplus \til P_{-1}^{-T} &\in [2^4] .
\nonumber
\end{align}
So only elements in $[1^4]$, $[2^4]$, $[3^2]$, $[4^2]$, and $[6^2]$ have lagrangian 1-eigenspaces.

\paragraph{Possible orders of elliptic-elliptic elements of Sp$_{\D}$(4,$\,\Z$).}

We can easily determine the characteristic polynomials of arbitrary powers of any $M\in[X]$: 
\begin{align}\label{powers}
[2^4]^2 &\subset  [1^4] \nonumber\\
[1^2 2^2]^2 &\subset  [1^4] \nonumber\\
[3^2]^3 &\subset  [1^4] \nonumber\\
[4^2]^2 &\subset  [2^4] & 
[4^2]^4 &\subset  [1^4] \nonumber\\
[6^2]^{2,4} &\subset  [3^2] &
[6^2]^3 &\subset  [2^4] &
[6^2]^6 &\subset  [1^4] \nonumber\\
[1^2 3]^3 &\subset  [1^4] \nonumber\\
[1^2 4]^2 &\subset  [1^2 2^2] &
[1^2 4]^4 &\subset  [1^4] \nonumber\\
[1^2 6]^{2,4} &\subset  [1^2 3] &
[1^2 6]^3 &\subset  [1^2 2^2] &
[1^2 6]^6 &\subset  [1^4] \nonumber\\
[2^2 3]^{2,4} &\subset  [1^2 3] &
[2^2 3]^3 &\subset  [1^2 2^2] &
[2^2 3]^6 &\subset  [1^4] \nonumber\\
[2^2 4]^2 &\subset  [1^2 2^2] &
[2^2 4]^4 &\subset  [1^4]
\\
[2^2 6]^{2,4} &\subset  [1^2 3] &
[2^2 6]^3 &\subset  [2^4] &
[2^2 6]^6 &\subset  [1^4] \nonumber\\
[3\cdot4]^{2,10} &\subset  [2^2 3] &
[3\cdot4]^{3,9} &\subset  [1^2 4] &
[3\cdot4]^{4,8} &\subset  [1^2 3] &
[3\cdot4]^{6} &\subset  [1^2 2^2] &
[3\cdot4]^{12} &\subset  [1^4] \nonumber\\
[3\cdot6]^{2,4} &\subset  [3^2] &
[3\cdot6]^{3} &\subset  [1^2 2^2] &
[3\cdot6]^{6} &\subset  [1^4] \nonumber\\
[4\cdot6]^{2,10} &\subset  [2^2 3] &
[4\cdot6]^{3,9} &\subset  [2^2 4] &
[4\cdot6]^{4,8} &\subset  [1^2 3] &
[4\cdot6]^{6} &\subset  [1^2 2^2] &
[4\cdot6]^{12} &\subset  [1^4] \nonumber\\
[5]^{5} &\subset  [1^4] \nonumber\\
[8]^{2,6} &\subset  [4^2] &
[8]^{4} &\subset  [2^4] &
[8]^{8} &\subset  [1^4] \nonumber\\
[10]^{2,4,6,8} &\subset  [5] &
[10]^{5} &\subset  [2^4] &
[10]^{10} &\subset  [1^4] \nonumber\\
[12]^{2,10} &\subset  [6^2] &
[12]^{3,9} &\subset  [4^2] &
[12]^{4,8} &\subset  [3^2] &
[12]^{6} &\subset  [2^4] &
[12]^{12} &\subset  [1^4]\, .\nonumber
\end{align}
Here we are using a notation where $[X]^a \subset [Y]$ means that $M^a\in[Y]$ if $M\in[X]$.  Note that $[X]^a \subset [Y]$ does \emph{not} imply that if $M_i\in[X]$ then $\prod_{i=1}^a M_i \in [Y]$:  this is only necessarily true if all $M_i$ are equal.

The smallest $N$ such that $[X]^N \subset [1^4]$ is $N=\text{lcm}(n_i)$ if $X = \prod n_i^{r_i}$.  The pattern of inclusions shown in \eqref{powers} repeats mod $N$ in the exponent.  Not shown in \eqref{powers} are all entries of the form $[X]^A \subset [X]$ which is true for all $A$ such that $\gcd(A,N)=1$.

For $M$ of type $(EE)$ --- i.e., no Jordan blocks of size greater than 1 --- then $[1^4] = \{ \I\}$ (and $[2^4] = \{ -\I\}$).  Then  the smallest $N$ such that $[X]^N\subset[1^4]$, which is shown as the last entry in each line of \eqref{powers}, is thus the order of unipotency of any $M\in[X]$.

\end{appendix}

\providecommand{\href}[2]{#2}\begingroup\raggedright\endgroup

\end{document}